\documentclass[aps,prd,preprint,superscriptaddress,nofootinbib,showkeys]{revtex4-1}
\usepackage{graphicx,amsmath,amssymb,bm,hyperref}

\newcommand{\eps}{\epsilon}
 \newcommand{\euv}{\epsilon_{\mathrm{UV}}}
\newcommand{\eir}{\epsilon_{\mathrm{IR}}}
\newcommand{\ms}{\Bigl(\frac{\mu^2 e^{\gamma_{\mathrm{E}}}}{4\pi}\Bigr)^{\epsilon}}
\newcommand{\dms}{\frac{(\mu^2 e^{\gamma_{\mathrm{E}}})^{\epsilon}}{\Gamma(1-\epsilon)}}
\newcommand{\dd}[1]{\frac{d^D {#1} }{(2\pi)^D}} 
 \newcommand{\fms}[1]{{#1}\!\!\!/}

\newcommand{\mc}{\mathcal}

\begin{document}
\title{Factorization of the dijet cross section in hadron-hadron collisions}

\def\KU{Department of Physics, Korea University, Seoul 02841, Korea} 
 
\author{Junegone Chay}
\email[E-mail:]{chay@korea.ac.kr}
\affiliation{\KU}
\author{Taewook Ha}
\email[E-mail:]{hahah@korea.ac.kr} 
\affiliation{\KU}
\author{Taehyun Kwon} 
\email[E-mail:]{aieamfirst@korea.ac.kr}
\affiliation{\KU}
 
\begin{abstract} \vspace{0.1cm}\baselineskip 3.0 ex 
The factorization theorem for the dijet cross section is presented in hadron-hadron collisions  
with a cone-type jet algorithm. We also apply the beam veto to the beam jets consisting of
the initial radiation.
The soft-collinear effective theory is employed to see the factorization structure transparently when there
are four distinct lightcone directions involved.
There are various types of divergences such as the ultraviolet and infrared divergences. And when the phase
space is divided to probe the collinear and the soft parts, there appears an additional divergence called rapidity
divergence. These divergences are sorted out and we will show that all the infrared and rapidity divergences 
cancel, and only the ultraviolet divergence remains. It is a vital step to justify the factorization. 
Among many partonic processes, 
we take $q\overline{q} \rightarrow gg$ as a specific example to consider the dijet cross section. 
The  hard and the soft functions have nontrivial color structure, while the jet and the beam functions 
are diagonal in operator basis. The dependence of the soft anomalous dimension on the jet 
algorithm and the beam veto is diagonal in operator space, and is cancelled by that of the jet and beam functions. 
We also compute the anomalous dimensions of the factorized components, and resum 
the large logarithms to next-to-leading logarithmic accuracy by solving the renormalization group equation.   
\end{abstract}
\maketitle

\baselineskip 3.0 ex 

\section{Introduction\label{intro}}
The study of jet physics in high energy scattering has reached a sophisticated level. Many of the 
factorization theorems for inclusive scattering processes have been established both in QCD 
and in soft-collinear effective theory (SCET)~\cite{Bauer:2000ew,Bauer:2000yr,Bauer:2001yt}. 
More differential quantities such as the transverse momentum dependence of the final-state 
particles or jets~\cite{Stewart:2013faa}, and the jet substructures~\cite{Dasgupta:2013ihk} 
have been studied. When we probe more differential quantities,  the factorization theorems 
should be first proved in order that each factorized part can be computed in perturbation 
theory offering predictive power.

In general, the factorization theorem states that the expression for physical observables is written as
the product or the convolution of the  hard, the collinear and the soft parts. In proving the 
factorization theorem, it is important to verify that each factorized part is infrared (IR)  finite. 
Otherwise the dependence of the renormalization scale does not solely come from the 
ultraviolet (UV) divergence, which invalidates the scaling behavior of the factorized parts. 
If some components are not IR finite in the factorized form, the factorized parts should be 
refactorized such that the redefined or rearranged quantities are IR finite. If the IR divergence 
still remains even after the rearrangement, the quantity at hand is not physical. 

A fully inclusive quantity is IR finite to all orders due to the 
Kinoshita-Lee-Nauenberg theorem~\cite{Kinoshita:1962ur,Lee:1964is} since the IR 
divergence from the virtual contribution is cancelled by that of the real contribution. 
It should hold true also for exclusive physical quantities such as the dijet cross section, 
even though the phase space for the real gluon emission is constrained by the jet algorithm 
and the beam veto. However, large logarithms appear due to the slight mismatch of the 
phase spaces for the virtual and real contributions.

The verification that each factorized part is IR finite in the dijet cross section from $e^+ e^-$ 
annihilation with various jet algorithms has been performed in  Refs.\cite{Chay:2015ila,Chay:2015dva}. 
On the other hand, here we take the dijet cross section in hadron-hadron collision with 
a cone-type jet algorithm \cite{Salam:2009jx} to establish the factorization theorem by carefully 
dissecting the phase space and performing the corresponding computation. This process is 
more complicated due to the complex color structure and the existence of the beam jets, 
thus more illuminating to show how to disentangle the interwoven structure of the dependence 
of the cross section on the jet algorithms and the beam veto. 

We employ SCET to present the 
factorization theorem for the dijet cross section because it is the appropriate effective theory 
to describe this process. The advantage of SCET is to establish the decoupling between the collinear
and the soft modes at the operator level, thus the factorization procedure manifest.
However, as far as the structure of the divergence is concerned,  there is
another type of divergence, called the rapidity divergence~\cite{Chiu:2011qc,Chiu:2012ir} in SCET.
It appears because we dissect the phase space into the collinear and the soft regions, and 
the soft region with small rapidity does not recognize the collinear region with large rapidity. 
In the full theory, there is no such divergence because there is no kinematic constraint. It is a 
good consistency check for the effective theory to see physical observables are free of the 
rapidity divergence. We show that the factorized parts do not have the rapidity divergence, 
though the individual contribution may possess one.  If a physical observable
is more differential, there may exist rapidity divergence in the collinear and soft parts separately, 
though they cancel in the total contribution. This rapidity divergence contributes to the 
additional renormalization group (RG) evolution. However, it is not our topic in this paper 
because there is no rapidity divergence in each factorized part in the inclusive jet cross section.

The dijet cross section in hadron-hadron collision is shown to be factorized into the hard, 
collinear and soft parts, which is schematically written as
\begin{equation} \label{schemesig}
\sigma_J \sim \mathrm{tr} (\mathbf{H} \otimes \mathbf{S}) \otimes B_{i/N_1} 
\otimes B_{j/N_2} \otimes \mathcal{J}_3 \otimes \mathcal{J}_4.
\end{equation}
A more rigorous expression will be derived in Sec.~\ref{fact}.  Here $\mathbf{H}$ is the 
hard function depending only on the hard scales, and $\mathbf{S}$ is the soft function 
which describes the soft radiations interspersed between the energetic particles.  The hard and soft 
functions are matrices in color space because they arise from different color channels. 
The incoming partons emit particles in the initial radiation, and the off-shell partons 
participate in the hard scattering. These processes are described by the beam functions 
$B_{i/N}$,  for the parton $i$ entering into the hard interaction from the hadron 
$N$~\cite{Stewart:2009yx}.  And $\mathcal{J}_i$ are the integrated jet functions 
describing the outgoing collinear particles in the final state, prescribed by a jet algorithm. 

The RG evolution of the functions in Eq.~\eqref{schemesig} resums a large set of logarithms. In providing
factorization, there is a hierarchy of scales. The hard scale is characterized by $Q$, the collinear scale is $Q\lambda$,
and the soft scale is $Q\lambda^2$, where $\lambda$ is a small parameter, appearing in SCET. Then at 
fixed order, there appears large logarithms with the ratios of these disparate scales, which may invalidate 
the perturbative series. Therefore these large logarithms should be resummed to all orders. 
In the dissected phase spaces, there appears only a singlet scale, and the resummation can be achieved by
solving the RG equation for each factorized part. However, we note that there are
other types of logarithms such as the logarithms of the small jet radius $R$~\cite{Alioli:2013hba, Kelley:2012zs} 
and nonglobal logarithms~\cite{Dasgupta:2001sh,Banfi:2002hw,Appleby:2003ai}, which are big challenges. 
Here we take the jet radius $R$ to be not too small ($R \sim 0.7$), and do not perform 
the small $R$-resummation. And we will not consider the nonglobal logarithms either because 
we are mainly interested in the factorization structure with four lightlike directions in the process. Last but not
least, we assume that the Glauber gluons, which are responsible for the interactions between the active partons
and the spectator partons, do not violate factorization.
 
 \begin{figure}[t] 
\begin{center}
\includegraphics[height=7cm]{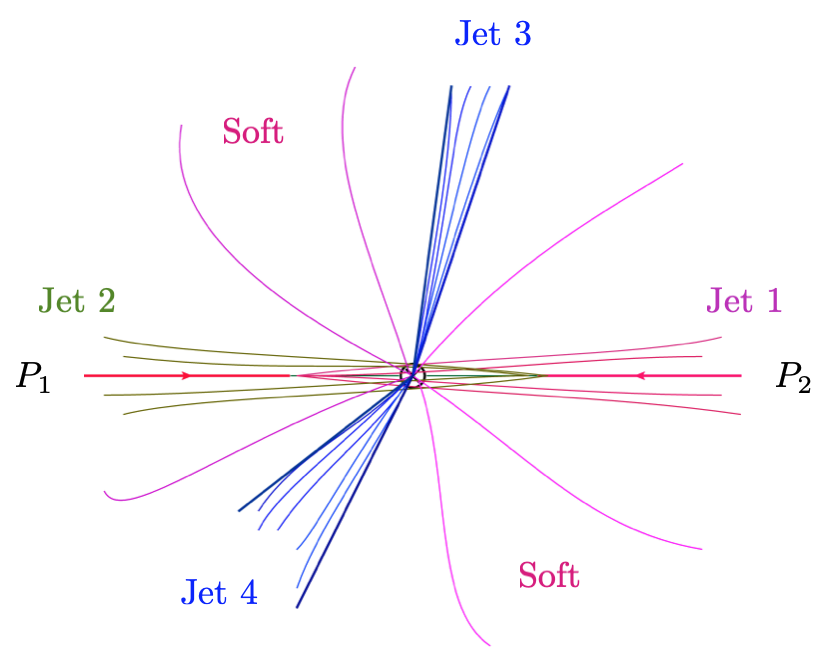}
\end{center}  
\vspace{-0.3cm}
\caption{\baselineskip 3.0ex A schematic diagram for the dijet production in hadron-hadron 
scattering. From the hadrons with momenta  $P_1$, and $P_2$, the partons, after emitting the initial-state
radiation, with the momenta 
$p_1$ and $p_2$ undergo a hard scattering.  Out of the hard scattering, the two jets with 
their momenta $p_3$ and $p_4$ are produced. The initial-state radiation produces beam jets. 
The soft momentum $p_s$ is either inside or 
outside the jets depending on the jet algorithm. \label{scheme}}
\end{figure}

The configuration of the dijet production is sketched in Fig.~\ref{scheme}. The incoming partons 
from the two hadrons with momenta $P_1$ and $P_2$ radiate the initial-state radiation, and finally 
have the momenta $p_1$ and $p_2$, and participate in the hard collision.  The remnant produces 
the beam jets in the beam directions. The outgoing particles form collimated beams of collinear particles, 
called jets, and the jets are clustered depending on the jet algorithms being employed. Jet algorithms 
in $e^+ e^-$ annihilation and in hadron-hadron scattering are different since the latter should preserve 
the boost invariance along the beam direction. However, the relations between the jet algorithms in 
the two cases are explained in Ref.~\cite{Hornig:2016ahz}, and we can employ the same 
cone-type jet algorithm with the rescaling of the jet radius accordingly. For central jets with
the rapidity close to zero, the cone-type jet algorithms in both cases take the same form.  
On the other hand, the initial-state radiation also forms beam jets in the beam direction. The beam 
jets are constrained by the rapidity cutoff $y_{\mathrm{cut}}$ to be contained in the beam direction, 
which will be discussed in detail. It is called the beam veto.
 
 In high-energy processes, there are hierarchies of scales involved at every stage of the scattering, 
 and they should be entangled to sort out the effect of the strong interaction. This is the essence in 
 the proof of the factorization, in which the hard, collinear and soft parts are factorized.
Furthermore it is important to guarantee that each factorized part contains only the UV divergence 
without any IR divergence. Another aspect is the rapidity divergence. In SCET, the collinear and 
the soft regions are separated by the rapidities. The collinear modes have large 
rapidity, while the soft modes have small rapidity. In computing the soft radiative correction, it 
reaches the large rapidity region, which cannot be recognized by the soft mode, and this causes 
the rapidity divergence. If the collinear and the soft modes do not have the same offshellness, 
the rapidity divergence vanishes in each sector. However, when they have the 
same offshellness, the rapidity divergence may remain in each sector though their sum vanishes. 
This affects the  RG evolution of the collinear and the soft parts.

Here we focus on the  dijet cross section to illustrate how carefully the UV and IR divergences 
along with the rapidity divergence should be treated. We examine various radiative corrections 
to next-to-leading order (NLO) and show that each factorized part is indeed IR, and rapidity finite. 
This type of analysis should be applied to other various differential processes for the rigorous 
proof of the factorization theorems.  Only after the remaining divergence is guaranteed to be 
of the UV origin, we can safely apply the RG equation to resum  
large logarithms. In this paper, we employ the pure dimensional regularization with the 
spacetime dimension $D=4-2\eps$ and the $\overline{\mathrm{MS}}$ scheme, 
in which we carefully distinguish the UV and the IR divergences in computing radiative 
corrections of the collinear and soft functions with a jet algorithm and a beam veto.  

The dijet cross section is described by $2\to 2$ processes at the parton level. We can analyze all the 
processes for phenomenology, but we rather choose the specific process $q\overline{q} \to gg$ 
to show how to treat all the types of the divergences consistently. This process involves the 
computation of the gluon jet function with the jet algorithm~\cite{Ellis:2010rwa}, and the quark beam 
function with the beam veto. 
Note that the quark jet function with the jet algorithm was calculated in 
Ref.~\cite{Chay:2015ila,Jouttenus:2009ns, Cheung:2009sg}. And the structure of the hard and 
soft functions is interesting and complicated enough to seek the consistency in the relations 
among the anomalous dimensions. 

This paper is organized as follows: The factorization of the dijet cross section in hadron-hadron 
collision is presented in Sec.~\ref{fact}. We take a specific example of the partonic process 
$q\overline{q} \rightarrow gg$ to express the individual factorized components explicitly.  
In Sec.~\ref{rap}, we briefly discuss the source of the rapidity divergence and explain how 
to introduce the rapidity regulator to treat the rapidity divergence. In Sec.~\ref{jetalg}, 
the jet algorithm is described to constrain the final-state particles to form jets. We also introduce 
the beam veto, which is analogous to the jet algorithm. But the beam veto is expressed in terms 
of the rapidity cutoff instead of the jet radius. In Sec.~\ref{collinearf}, the collinear functions are 
computed at NLO.  In Sec.~\ref{gjetf}, the gluon jet function and its anomalous dimensions 
are computed with the cone-type algorithm at NLO, and in Sec.~\ref{cdfn}, the quark beam functions  
and its anomalous dimension are computed. The soft function is computed in Sec.~\ref{sofun}, in which
the effect of the jet algorithm and the beam veto are implemented. In Sec.~\ref{evol}, we solve 
the RG equations for the factorized functions and resum the large logarithms to next-to-leading 
logarithmic (NLL) order. And the independence of the renormalization scale in the dijet cross 
section is confirmed explicitly. In Sec.~\ref{conc}, the conclusion and the perspective are presented.  
 
\section{Factorization of the dijet cross section\label{fact}}
We consider the dijet  cross section in hadron-hadron collisions 
\begin{equation}
N_1 (P_1) + N_2 (P_2) \rightarrow J_3 (p_3) + J_4 (p_4) + X,
\end{equation}
where $N_1$ and $N_2$ are incoming hadrons (protons in the case of LHC), $J_3$ and $J_4$ denote 
two energetic collinear jets and $X$ represents all the other particles. The momenta of the incoming 
partons $p_1$ and $p_2$ can be written in terms of the hadronic momenta $P_1$ and $P_2$ as 
$p_1 = z_1 P_1$ and $p_2 = z_2 P_2$ respectively, where $z_i$ refer to the longitudinal momentum 
fractions.  We choose the beam directions to be in the $n_1$ and $n_2$ lightlike directions with 
$n_1^2 = n_2^2=0$,  $n_1 \cdot n_2 =2$, and $\overline{n}_1 = n_2$. For convenience, we 
choose the beam directions in the $z$ direction as $n_1^{\mu} = (1,0,0,1)$, $n_2^{\mu} = (1, 0, 0, -1)$. 
The jet directions are chosen to be in the lightlike directions $n_3$, and $n_4$. And we consider the 
dijets away from the beam direction, which can be stated as 
$n_1 \cdot n_3 \sim n_1 \cdot n_4\sim \mathcal{O}(1)$.

The dijet cross section  in SCET is written as
\begin{eqnarray}
\sigma (N_1 N_2 \rightarrow J_3 J_4 X) &=& \frac{1}{2S} \prod_{X_3, X_4, X} (2\pi)^4 
\delta^{(4)} (P_1^{\mu} +P_2^{\mu} -p_3^{\mu}
-p_4^{\mu} -p_{X}^{\mu}) \nonumber \\
&\times& \sum_{IJ} C_I C_J^* \langle  N_1 N_2 |O_J^{\dagger}  |X_3, X_4, X \rangle 
\langle X_3, X_4, X|O_I|N_1 N_2 \rangle.
\end{eqnarray}
Here $\prod_{X_3, X_4, X}$ denotes the phase space for the final-state particles, and $S$ is the hadronic 
center-of-mass energy squared. The set of operators $O_I$ are the SCET operators for 
$2\rightarrow 2$ processes and $C_I$ are the Wilson coefficients obtained by matching SCET and 
full QCD \cite{Ellis:1985er,Kelley:2010fn}.
 
In SCET, the collinear momentum $p^{\mu}$ in the lightlike $n$ direction can be decomposed as
\begin{equation}
p^{\mu} = \overline{n}\cdot p \frac{n^{\mu}}{2} + p_{\perp}^{\mu} + n\cdot p \frac{\overline{n}^{\mu}}{2}
= \mathcal{P}^{\mu} + n\cdot p \frac{\overline{n}^{\mu}}{2}
\end{equation}
where it scales as $p^{\mu} = (\overline{n}\cdot p, p_{\perp}, n\cdot p) = (p^- , p_{\perp}, p^+) \sim
Q(1, \lambda, \lambda^2)$. Here $Q$ is the largest component of the momentum and $\lambda$ is the small
parameter in SCET. The momentum components of order $Q$ and $Q\lambda$ are called the label momenta,
and they are integrated out, and the remaining degrees of freedom are of order $Q\lambda^2$.
 
The operators can be categorized by the partons participating in the hard scattering processes. 
If the initial- and final-state particles consist of quarks or antiquarks, the set of the SCET 
collinear operators for $q\overline{q} \rightarrow q' \overline{q'}$ are given as
\begin{equation}
\mathcal{O}_1  =  \overline{\chi}_2  T^a \gamma^{\mu}  \chi_1 \cdot 
\overline{\chi}_4   T^a \gamma_{\mu}   \chi_3, \ \
\mathcal{O}_2  =  \overline{\chi}_2  \gamma^{\mu}   \chi_1 \cdot 
\overline{\chi}_4  \gamma_{\mu}   \chi_3,
\end{equation}
where the collinear fields $\chi = W^{\dagger}\xi$ are the collinear gauge-invariant combination 
with the collinear Wilson line $W$
\begin{equation} \label{cowil}
W= \sum_{\mathrm{perm}} \exp \Bigl[ -g \frac{\overline{n}\cdot A_n}{\overline{n} 
\cdot \mathcal{P}}\Bigr],
\end{equation}
for the collinear gauge field $A_n$ in the lightlike $n$ direction. The SU(3) generators $T^a$ for the strong 
interaction are in the fundamental representation. These operators are responsible for the 
scattering of $q \overline{q} \rightarrow q\overline{q}$, $qq \rightarrow qq$, 
$\overline{q}\overline{q}\rightarrow \overline{q} \overline{q}$ including different types 
of quarks, which are related by the appropriate crossing symmetry.

For the processes $q\overline{q} \rightarrow gg$, $gg \rightarrow q \overline{q}$, 
$gq\rightarrow gq$ and $g\overline{q} \rightarrow g\overline{q}$,
the relevant SCET collinear operators are given by\footnote{\baselineskip 3.0ex 
In Ref.~\cite{Chiu:2009mg}, another independent set of operators are introduced:
$\tilde{\mathcal{O}}_1 = \overline{\chi}_2  \chi_1 B_{\perp 4}^{\mu a} B_{\perp \mu 3}^{a}$, 
$\tilde{\mathcal{O}}_2 = d_{abc}\overline{\chi}_2 T_c\chi_1 B_{\perp 4}^{\mu a} 
B_{\perp \mu 3}^{b} $, and $\tilde{\mathcal{O}}_3 = if_{abc}\overline{\chi}_2 T_c\chi_1 
B_{\perp 4}^{\mu a} B_{\perp \mu 3}^{b}$. They are related by 
$\tilde{\mathcal{O}}_1 = \mathcal{O}_3$, 
$\tilde{\mathcal{O}}_2 = \mathcal{O}_1 + \mathcal{O}_2 - \mathcal{O}_3/N$, and 
$\mathcal{O}_3 = \mathcal{O}_1 -\mathcal{O}_2$.} 
\begin{equation}
\mathcal{O}_1 = \overline{\chi}_2 T_a T_b \chi_1 B_{\perp 4}^{\mu a} B_{\perp \mu 3}^{b}, \
\mathcal{O}_2 = \overline{\chi}_2 T_b T_a \chi_1 B_{\perp 4}^{\mu a} B_{\perp \mu 3}^{b}, \  
\mathcal{O}_3 = \overline{\chi}_2  \chi_1 B_{\perp 4}^{\mu a} B_{\perp \mu 3}^{a},
\end{equation}
where $B_{\perp i}^{\mu} = [ W_i^{\dagger} iD_{i}^{\perp\mu} W_i]/g$ is the collinear gauge-invariant 
gluon field in the $n_i$ direction in SCET.
For the process $gg\rightarrow gg$, there are 9 independent collinear SCET operators, 
which are of the form $\mathcal{O}_i =  T_i^{abcd}
B_{\perp 2}^{\pm a} B_{\perp  1}^{\pm b} B_{\perp 4}^{\pm c} B_{\perp   3}^{\pm d}$ ($i=1,\cdots, 9$), 
where $\pm$ indicates
the helicity of the gluons. The explicit form of the operators can be found in Ref.~\cite{Kelley:2010fn}. 

The factorization procedure can be performed for any partonic processes, but it is illustrative to pick 
up a single partonic process and treat 
the factorization in detail.  As a specific example, we consider the partonic process 
$q \overline{q} \rightarrow gg$. The relevant operators with the redefinition of the collinear 
fields to decouple the soft interaction $\chi \rightarrow Y\chi$, 
$B_{\perp\mu}^a \rightarrow \mathcal{Y}^{ab} B_{\perp\mu}^b$ are given by
\begin{equation}
O_I = \Bigl(\overline{\chi}_2^{\alpha}  \chi_1^{\beta}   B_{4\perp\mu}^a B_{3\perp }^{b\mu}\Bigr)
\Bigl( Y_2^{\dagger}\mathcal{Y}_4^{\dagger aa'} T_I^{a'b'} 
\mathcal{Y}_3 ^{b'b} Y_1\Bigr)_{\alpha\beta}, \label{qqgg}
\end{equation}
where $T_1^{ab} = T^a T^b$, $T_2^{ab} = T^b T^a$ and $T_3^{ab} = \delta^{ab}$. 
The indices $a$, $b$ ($\alpha$, $\beta$) refer to those of the adjoint (fundamental) representation. 
The soft Wilson line $Y_i$ associated with the $n_i$-collinear fermion is given in the fundamental 
representation, while the soft Wilson line $\mathcal{Y}_i$ from the $n_i$-collinear 
gluon is given in the adjoint representation:
\begin{equation} \label{sowil}
Y_i= \sum_{\mathrm{perm}} \exp \Bigl[ -g\frac{n_i\cdot A_s^{\alpha} 
T_{\alpha}}{n_i\cdot \mathcal{P}}\Bigr], \ \ 
\mathcal{Y}_i = \sum_{\mathrm{perm}} \exp \Bigl[ -g\frac{n_i\cdot A_s^{a} 
T_{a}}{n_i\cdot \mathcal{P}}\Bigr],
\end{equation}
where $T_{\alpha}$ ($T_a$) are the generators of the color SU(3) in the fundamental (adjoint) 
representations. After this redefinition, the collinear modes and the soft modes are decoupled. 

The collinear matrix element for the operators in Eq.~(\ref{qqgg}) is given by
\begin{equation}
\sum_{X_3,X_4} \langle N_1 N_2|\overline{\chi}_1^{\rho}  \chi_2^{\sigma} B_{\perp 3}^{c\nu} 
B_{\perp 4 \nu}^d|X_3 X_4\rangle \langle X_3 X_4| \overline{\chi}_2^{\alpha} \chi_1^{\beta} 
B_{\perp 4 \mu}^a B_{\perp 3 }^{b\mu} |N_1 N_2\rangle,
\end{equation}
and it can be expressed in terms of the gluon jet functions for the final-state particles and the beam 
functions for the initial-state particles. The gluon jet functions in the $n_3$ and $n_4$ directions 
are defined as
\begin{eqnarray}
\sum_{X_3} \langle 0|B_{\perp 3}^{c\nu}  |X_3\rangle \Theta_J \langle X_3|
B_{\perp 3 }^{b\mu}  |0\rangle 
&=& -g_{\perp}^{\mu\nu} \delta^{bc}\int \frac{d^4p_3}{(2\pi)^3}  J_g (p_3^2), \nonumber \\
\sum_{X_4} \langle 0| B_{\perp 4}^{d\nu} |X_4\rangle \Theta_J \langle X_4|B_{\perp 4 }^{a\mu}
 |0\rangle 
&=& -g_{\perp}^{\mu\nu} \delta^{ad}\int \frac{d^4p_4}{(2\pi)^3}    J_g (p_4^2),
\end{eqnarray}
where $\Theta_J$ denotes the jet algorithm to be employed.  The jet functions are normalized to 
$\delta (p_i^2)$ at tree level.

The quark beam functions are defined as
\begin{equation} \label{beamf}
B_{q/N} (t, z= \omega/p^-, \mu) =  \langle N (P)| \theta (\omega) \overline{\chi}_n  
\delta (t + \omega \mathcal{P}^+ ) 
\frac{\fms{\overline{n}}}{2} \Theta_B [\delta (\omega -\mathcal{P}) \chi_n]| N(P)\rangle.
\end{equation}
 The beam veto is 
denoted as $\Theta_B$, and will be described in detail. The beam function is normalized 
as $\delta (t)\delta (1-z)$ at tree level. The unmeasured beam function is given by
\begin{equation} \label{unbeam}
B_{q/N}(z, \omega \delta,\mu) = \int dt B_{q/N}(t,z,\mu).
\end{equation}
In integrating over $t$, the dependence on the beam veto enters into the integrated beam function, and 
$\delta$ always appears in a combination $\omega \delta$ in the beam function.

Then the  dijet cross section is factorized as\footnote{To be rigorous, the effect of the Glauber gluons 
should be implemented to prove factorization.}
\begin{align}
\sigma = \frac{1}{64\pi N_c^2} &\sum_{IJ} \int \frac{dt}{s^2} H_{IJ} (\mu)  S_{JI} (\mu) 
\int   dz_1   B_{q/N_1}( z_1, \omega_1\delta_1,\mu)
\int  dz_2  B_{\bar{q}/N_2}(z_2, \omega_2 \delta_2, \mu) \nonumber \\
&\times \mathcal{J}_{g3} (\omega_3 \delta_3,  \mu)\mathcal{J}_{g4}  (\omega_4 \delta_4,\mu) 
+(q \leftrightarrow \overline{q}), 
\end{align}
where $N_c$ is the number of colors, and  $s$ and $t$ are the partonic Mandelstam variables.
The soft function $S_{JI}$ is defined as 
\begin{equation} \label{softfunction}
S_{JI} = \sum_{X_s}  \mathrm{tr} \langle 0| Y_1^{\dagger} (\mathcal{Y}_3^{\dagger} 
T_J^{\dagger} \mathcal{Y}_4)^{ba} Y_2 |X_s\rangle \Theta_J \Theta_B
\langle X_s |Y_2^{\dagger} (\mathcal{Y}_4^{\dagger} T_I \mathcal{Y}_3)^{ab} Y_1|0\rangle,
\end{equation}
and the dependence on the cone sizes $\delta_i$ is suppressed. 
The integrated jet function $\mathcal{J}_{gi}$ is defined as
\begin{equation}
\mathcal{J}_{gi} (\omega_i \delta_i, \mu)= \int dp_i^2 J_g (p_i^2, \omega_i \delta_i,\mu), \ (i=3,4).
\end{equation}
The dependence on the cone sizes $\delta_i$ come from the jet algorithm and the beam veto. In the final 
expression, we will use the same size of the jet radius $\delta_3 = \delta _4 = \delta$, and $\delta_1 =
\delta _2 = e^{-y_{\mathrm{cut}}}$. The purpose of distinguishing the jet radii in the 
intermediate step is to see explicitly how the anomalous dimensions in each collinear part are combined
to cancel the total anomalous dimensions when those of the hard and the soft parts are added.

If we are interested in the dijet invariant mass distribution $m_{j3}^2 = (p_3+l)^2$, $m_{j4}^2 = (p_4 +l)^2$,
the differential cross section with respect to the invariant jet masses is given by
\begin{align}
&\frac{d\sigma}{dm_{j3}^2 dm_{j4}^2} = \frac{1}{64\pi N_c^2}  \sum_{IJ} \int \frac{dt}{s^2}H_{IJ} 
\int  dz_1   B_{q/N_1}(z_1, \omega_1 \delta_1, \mu) \int  
dz_2  B_{\bar{q}/N_2}(z_2, \omega_2 \delta_2, \mu)  \nonumber \\
&\times \int dl_+ dl_-  \tilde{S}_{JI}(l_+, l_-) J_g (m_{j3}^2 - \overline{n}\cdot p_3 l_-, \omega_3\delta_3,\mu) 
J_g (m_{j4}^2 -\overline{n}_4 \cdot  p_4 l_+, \omega_4 \delta_4, \mu), 
\end{align}
with $l_+ = n_3\cdot l$ and $l_- = n_4\cdot l$. The differential soft function $\tilde{S}_{JI} (l_+, l_-) $ 
is defined as
\begin{equation}
\tilde{S}_{JI} (l_+, l_-) =    \mathrm{tr} \langle 0| Y_1^{\dagger} (\mathcal{Y}_3^{\dagger} T_J^{\dagger} 
\mathcal{Y}_4)^{ba} Y_2   \delta (l_+ + n_3 \cdot \mathcal{P}_s) \Theta_J  \Theta_B
\delta (l_- + n_4  \cdot \mathcal{P}_s)
 Y_2^{\dagger} (\mathcal{Y}_4^{\dagger} T_I \mathcal{Y}_3)^{ab} Y_1|0\rangle.
\end{equation}

\section{Treatment of the rapidity divergence\label{rap}}
 The UV and IR divergences are handled by the dimensional regularization, in which they are 
 expressed as poles in $\euv$ and $\eir$  respectively with the $\overline{\mathrm{MS}}$ scheme. 
 However, a new type of divergence, called the rapidity divergence,  shows up because the 
 phase space is divided into the collinear and soft regions. If the collinear and the soft 
 modes have the same magnitude of the invariant mass, they are characterized by their 
 rapidities. The rapidity divergence arises because the soft modes with small rapidity
 cannot recognize the collinear region with large rapidity. If we write the $n$-collinear 
 momentum as  $k^{\mu} =(\overline{n}\cdot k, k_{\perp}, n\cdot k) =(k^- , k_{\perp}, k^+ )$ 
 with the lightcone vectors satisfying  $n^2 = \overline{n}^2 =0$ and $n\cdot \overline{n}=2$, 
 the rapidity divergence occurs in the phase space where $k^-$ approaches infinity, 
 while $\mathbf{k}_{\perp}^2$ is fixed. In the same spirit of the dimensional regularization, 
 we modify the region with large rapidity to extract the rapidity divergence. 
 
In extracting the rapidity divergence, we have to be careful in distinguishing the spurious 
divergence coming from the region $k^-\rightarrow 0$ from the real rapidity divergence.  
In the real contribution, because $k^-$ is bounded from above, the rapidity divergence as 
$k^- \rightarrow \infty$ does not arise. However, there appears the divergence as $k^-$ 
approaches zero. This unwanted 
divergence is removed by the zero-bin subtraction~\cite{Chiu:2011qc, Chiu:2012ir},
 which removes the soft limit in the collinear region to avoid double counting. Since 
 the zero-bin  contribution is the same as  the soft contribution except its sign, the rapidity 
 divergence is cancelled when the contributions of the collinear and the  soft sectors are 
 combined. One of the explicit examples is presented in Ref.~\cite{Chay:2021arz}.
It is consistent with QCD in which there is no rapidity divergence since there is no kinematic 
separation. When the collinear and the soft modes have different offshellness, 
there is no rapidity divergence in each sector. The  dijet cross section we consider here 
belongs to this case, and we verify this explicitly in this paper.
 
In order to regulate the rapidity divergence, one of the authors has constructed consistent rapidity 
regulators for the collinear and the soft sectors~\cite{Chay:2020jzn}. The basic idea is to modify 
the collinear Wilson line by attaching the regulator of the form $(\nu/\overline{n}\cdot k)^{\eta}$
for the $n$-collinear field, where the rapidity divergence arises for $\overline{n}\cdot k \rightarrow \infty$. 
This prescription was originally proposed in Refs.~\cite{Chiu:2011qc,Chiu:2012ir} , and use the same regulator
as far as the collinear rapidity regulator is concerned. However, we require that the rapidity regulator 
for the soft sector come from the same source as the collinear radiation because 
the small rapidity limit for the soft gluons should come from the collinear radiations. Therefore the soft rapidity
regulator should take the same form $(\nu/\overline{n} \cdot k)^{\eta}$ as the collinear rapidity regulator. 
But as we can see from 
Eqs.~\eqref{cowil} and \eqref{sowil}, $\overline{n}\cdot \mathcal{P}$ and $n\cdot \mathcal{P}$ 
are employed, and we write the soft rapidity
regulator conforming to the expression in the soft Wilson line. 

Let us consider a collinear current 
$\chi_{n_1} W_{n_1} S_{n_1}^{\dagger} \Gamma S_{n_2} W_{n_2}^{\dagger} \chi_{n_2}$
as an example of constructing the rapidity regulators. By inserting the collinear and the soft 
Wilson lines $W_{n_i}$ and $S_{n_i}$, the current is collinear and soft gauge invariant. For the 
collinear Wilson line $W_{n_1}$ and the soft Wilson line $S_{n_2}$, the Wilson lines in 
Eqs.~\eqref{cowil} and \eqref{sowil} are modified with the rapidity regulators as
\begin{align} \label{rareg} 
W_{n_1} &= \sum_{\mathrm{perm.}} \exp \Bigl[
-\frac{g}{\overline{n}_1\cdot \mathcal{P}} \Bigl(\frac{\nu}{|\overline{n}_1\cdot
\mathcal{P}|}\Bigr)^{\eta} \overline{n}_1 \cdot A_{n_1}\Bigr], \nonumber \\ S_{n_2} 
&=\sum_{\mathrm{perm.}} \exp\Bigl[- \frac{g}{n_2\cdot \mathcal{P}}
\Bigl(\frac{\nu}{|n_2\cdot \mc{P}|} \frac{n_1\cdot n_2}{2}\Bigr)^{\eta} n_2\cdot A_s
\Bigr], 
\end{align} 
where $\mathcal{P}$ is the operator extracting the momentum. The collinear Wilson line is obtained by 
integrating out the offshell modes when the $n_1$-collinear gluons are emitted from the $n_2$-collinear particle. 
The soft Wilson line $S_{n_2}$ is obtained by exponentiating the emitted soft gluons from the $n_2$ collinear particle
to all orders. In regulating the rapidity divergence, we take the limit 
$\overline{n}_1 \cdot k\rightarrow \infty$ for these soft gluons and the momentum can be written as 
$k^{\mu} \approx (\overline{n}_1\cdot k) n_1^{\mu}/2$. Then we can write $n_2 \cdot k 
\approx  (\overline{n}_1\cdot k) n_1\cdot n_2/2$. Therefore the rapidity regulator can be written in the form
\begin{equation}
\Bigl(\frac{\nu}{\overline{n}_1 \cdot k}\Bigr)^{\eta} \xrightarrow[\bar{n}_1\cdot k \to \infty]{}
\Bigl(\frac{\nu}{n_2\cdot k} \frac{n_1\cdot n_2}{2}\Bigr)^{\eta}.
\end{equation}
This is different from the soft regulator suggested in Refs.~\cite{Chiu:2011qc,Chiu:2012ir}, because they 
considered only back-to-back currents. We emphasize that the directional dependence $n_1\cdot n_2$ is 
important to compute the soft function along with its anomalous dimensions.   
The remaining Wilson lines 
can be obtained by switching $n_1$ and $n_2$. The point in selecting the rapidity regulator 
is to trace the same emitted gluons both in the collinear and the soft sectors, which are 
eikonalized to produce the Wilson lines.  

\section{Jet algorithm and beam veto\label{jetalg}}
 
At the parton level, the dijet production from hadron-hadron scattering and the process 
$e^+ e^-\rightarrow 4$ jets are similar since they are related by the crossing symmetry.  
But in hadron-hadron scattering, only the final-state partons are organized by the jet algorithm, 
while all the final-state partons in $e^+ e^-\rightarrow 4$ jets are scrutinized by the jet algorithm. 
This affects the soft function and its anomalous dimension, which depend on the jet cone size. However,
it turns out that the anomalous dimension of the soft function  depending on the jet cone size 
is diagonal in color basis, which cancels the cone size dependence of the anomalous dimension 
in the jet function. 
 
We consider the cone-type jet algorithm at NLO, in which there are at most two particles inside a jet. 
At this order, we choose the jet axis in the $n$ direction. The jet axis may be chosen as the thrust axis, or 
the weighted average of the rapidity and the azimuthal angle 
over the transverse energy. Then the particles inside a jet should satisfy the condition 
$\theta_i <R$. Here $\theta_i$ is the angle of the $i$-th particle with respect to the jet axis, 
and $R$ is the jet cone size.

The jet algorithm can be expressed in terms of the lightcone momenta 
as follows~\cite{Chay:2015ila,Cheung:2009sg,Ellis:2010rwa}:
\begin{align}  
& \frac{n_3 \cdot l}{\overline{n}_3\cdot l} < \delta^2, \ \ n_3 \ \mathrm{jet}, \label{n3jet} \\
& \frac{n_4 \cdot l}{\overline{n}_4\cdot l} < \delta^2, \ \ n_4 \ \mathrm{jet},   \label{n4jet}\\
& l_0 < \Lambda, \ \ \mathrm{jet \ veto},   \label{jetveto}
\end{align}
where $\delta = \tan (R/2)$. If particles in the $n_3$ ($n_4$) directions should belong to 
the $n_3$ jet ($n_4$-jet), their lightcone momenta should satisfy Eq.~\eqref{n3jet} [Eq.~\eqref{n4jet}].
The jet veto applies to soft particles. If the energy of the soft particle gets larger than some veto scale 
$\Lambda$, and if they are outside the jet cones specified by $n_3$ or $n_4$, they should be vetoed. 
Expressing the jet veto in an equivalent way, the energy of the soft particles should be smaller than $\Lambda$
everywhere. 

This jet algorithm is employed in $e^+ e^-$ 
collisions, but we can retain this form with the understanding that $R$ is actually replaced 
by $\mathcal{R}/\cosh y_J$, where $\mathcal{R}$ is the cone size in the 
pseudorapidity-azimuthal angle space, and $y_J$ is the pseudorapidity of the jet 
in hadron-hadron scattering \cite{Hornig:2016ahz}. For the jet veto, the transverse
momentum cutoff $p_T<p_T^{\mathrm{cut}}$ is replaced by 
$E<p_T^{\mathrm{cut}} \cosh y_J =\Lambda$, where $\Lambda$ is the energy 
outside the jets acting as a jet veto.  The jet veto is needed to guarantee that the final 
states form a dijet event.  Eqs.~\eqref{n3jet}, \eqref{n4jet} and \eqref{jetveto} as a whole 
incorporate the jet algorithm. But when there is no confusion, we sometimes refer to the first two equations 
as the jet algorithm since they are the conditions for the particles to 
be inside the jet, and refer to the third equation as the jet veto. 

The beam jets, described by the beam function in Eq.~\eqref{beamf}, should also be 
confined in the beam directions, which could be prescribed following the jet algorithm 
for the final-state particles. The beam veto $\Theta_B$ at NLO can be expressed as
\begin{equation}  \label{beamveto}
\Theta_B =  \Theta \Bigl( \frac{n_i\cdot l}{\overline{n}_i\cdot l} <\delta^2  \Bigr) = 
\Theta \Bigl( \frac{n_i\cdot l}{\overline{n}_i\cdot l} <e^{-2y_{\mathrm{cut}}}  \Bigr),
\ \ (i=1,2).
\end{equation} 
Because there is only a single particle in the beam function at NLO, Eq.~\eqref{beamveto} is enough.

We apply the beam veto with the center-of-mass energy 
$E_{\mathrm{cm}}$ and a rapidity cut of $y_{\mathrm{cut}}$. Then the beam function can 
be written as
\begin{equation}
B_i (z_i, \mu) = B_i (E_{\mathrm{cm}}, y_{\mathrm{cut}}, z_i, \mu) = B_i (z_i E_{\mathrm{cm}}
e^{-y_{\mathrm{cut}}}, z_i ,\mu),
\end{equation}
which means that the beam function depends on $E_{\mathrm{cm}}$ and $y_{\mathrm{cut}}$, 
always in the combination $E_{\mathrm{cm}} e^{-y_{\mathrm{cut}}}$. Here $z_i$ is the longitudinal
momentum fraction of the parton. From now on, we apply the beam veto to 
the beam functions, and for the beam we use the relation~\cite{Hornig:2016ahz}
\begin{equation}
\omega_i \delta_i = \omega_i \tan \frac{R_i}{2} \rightarrow z_i E_{\mathrm{cm}} e^{-y_{\mathrm{cut}}}.
\end{equation}

The soft function is influenced both by the jet algorithm and the beam veto. The dependence is 
expressed in terms of the jet size $\delta =\tan R/2$, but it should be understood that 
the beam veto can be expressed as the rapidity cutoff by replacing $\delta = e^{-y_{\mathrm{cut}}}$.

For power counting in SCET, the $n$-collinear momentum scales as $p_n^{\mu} = 
(\overline{n}\cdot p, p_{\perp}, n\cdot p) \sim Q (1, \lambda, \lambda^2)$, where $\lambda$ is 
the small parameter. Then the ultrasoft (usoft) momentum scales 
as $p_s^{\mu} \sim Q(\lambda^2, \lambda^2,\lambda^2)$. And we also take 
$\delta\sim \mathcal{O}(\lambda)$ and $\Lambda \sim \mathcal{O}(Q\lambda^2)$
for definiteness. We may need other degrees of freedom if we are interested in the small $R$ 
resummation~\cite{Chien:2015cka}. But this topic is beyond the scope of the paper.
 
\section{Collinear functions\label{collinearf}}
\subsection{Gluon jet function\label{gjetf}}
\begin{figure}[t] 
\begin{center}
\includegraphics[width=4cm]{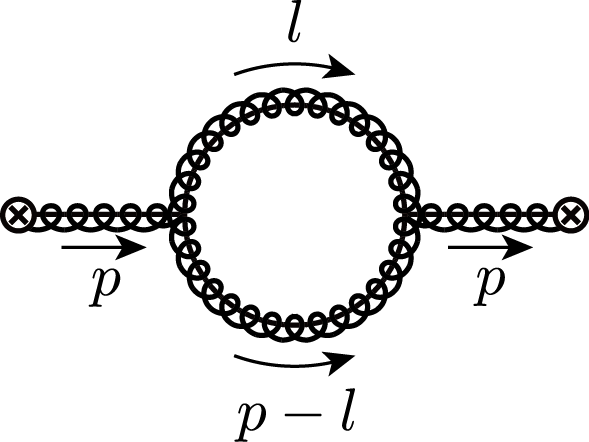}
\end{center}  
\vspace{-0.5cm}
\caption{\baselineskip 3.0ex   Assignment of the momenta for the final-state gluons. \label{moflow}}
\end{figure}

\begin{figure}[t] 
\begin{center}
\includegraphics[width=12cm]{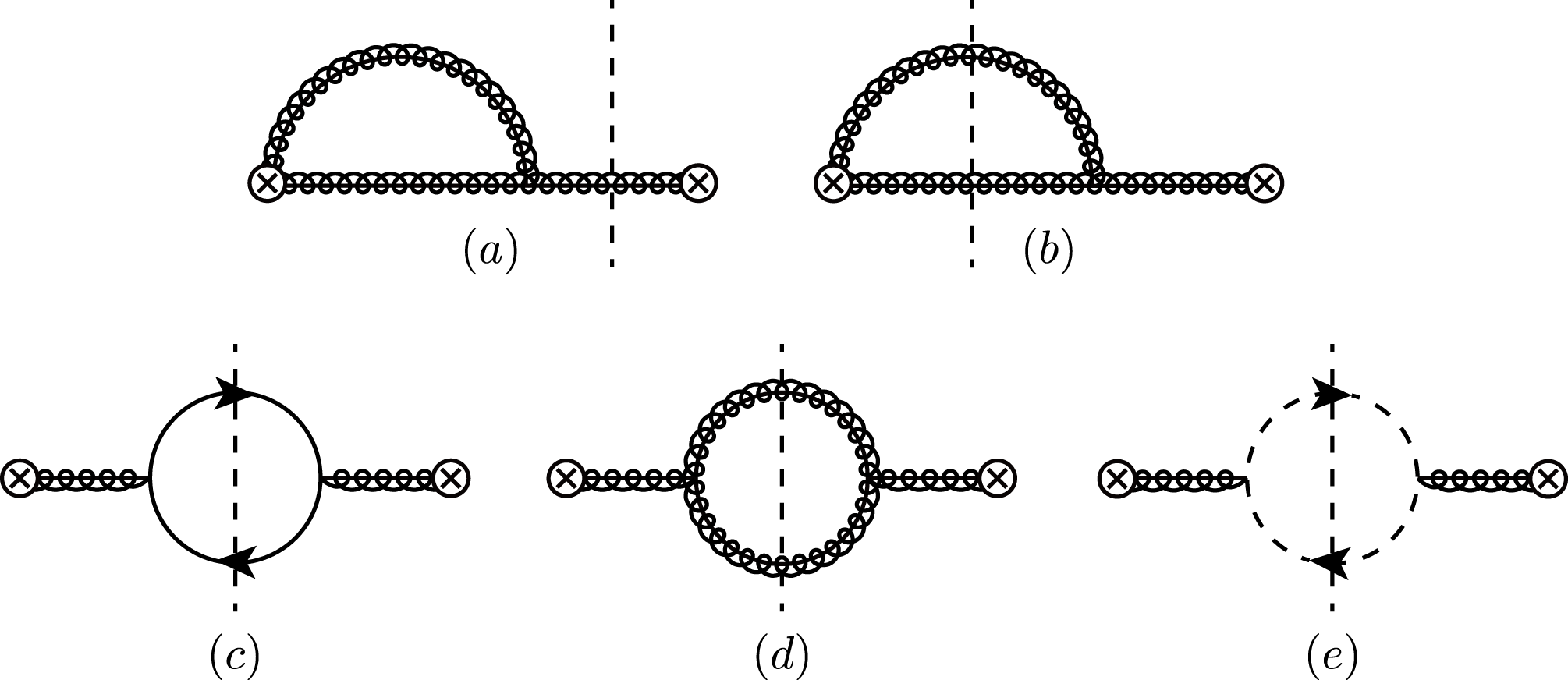}
\end{center}  
\vspace{-0.5cm}
\caption{\baselineskip 3.0ex   Feynman diagrams for the gluon jet function. Curly lines are gluons, 
solid lines with arrows are fermions, dashed lines with arrows are ghost particles. The vertical dashed 
lines represent the unitarity cut. (a)  Virtual correction  (b) real gluon emission,  and the mirror images 
are omitted. The remainder represents the contributions from the cuts of (c) a fermion loop (d) a gluon 
loop and (e) a ghost loop. The diagrams for the wave function renormalization are not shown 
here. \label{gluejet}}
\end{figure}

The inclusive gluon jet function has been computed to one-loop order \cite{Becher:2009th}, and 
two-loop order \cite{Becher:2010pd} without any jet algorithms. Here we compute the gluon 
jet function with the cone jet algorithm at one loop. The cone jet algorithm at NLO
involves at most two particles inside a jet. In computing the jet function, the Feynman diagrams for 
the matrix element squared is schematically shown in Fig.~\ref{moflow}. The loop includes other particles. 
(See Fig.~\ref{gluejet}.) If we make a unitarity cut in any of the internal lines, the cut lines 
correspond to the final-state particles. For example,  if a single leg is cut, it represents a 
single final-state particle with the virtual correction. If the loop is cut, it represents 
two final-state particles, to which the jet algorithm is applied.

The momentum of the jet is given by $p$, and the momenta of the two gluons are labeled as $l$ and $p-l$ 
respectively. Suppose that the jet is collinear in the $n$ lightcone direction. Then the momenta 
of the gluons can be written as
\begin{equation}
p_1 =  (l_-, l_{\perp}, l_+), \ p_2 =(\omega-l_-, -l_{\perp}, p^2 /\omega -l_+), 
\end{equation}
where $\omega= \overline{n}\cdot p = p_-$.  The energies of the gluons and the invariant mass 
squared of the jet are given by
\begin{equation}
E_1 = \frac{1}{2} (l_- +l_+ ), \ \ E_2 = \frac{1}{2} (\omega-l_- +\frac{p^2}{\omega}-l_+), 
\ p^2 =\frac{\omega l_+}{1-l_-/\omega}.
\end{equation}

\begin{figure}[b]
\begin{center}
\includegraphics[height=5.3cm]{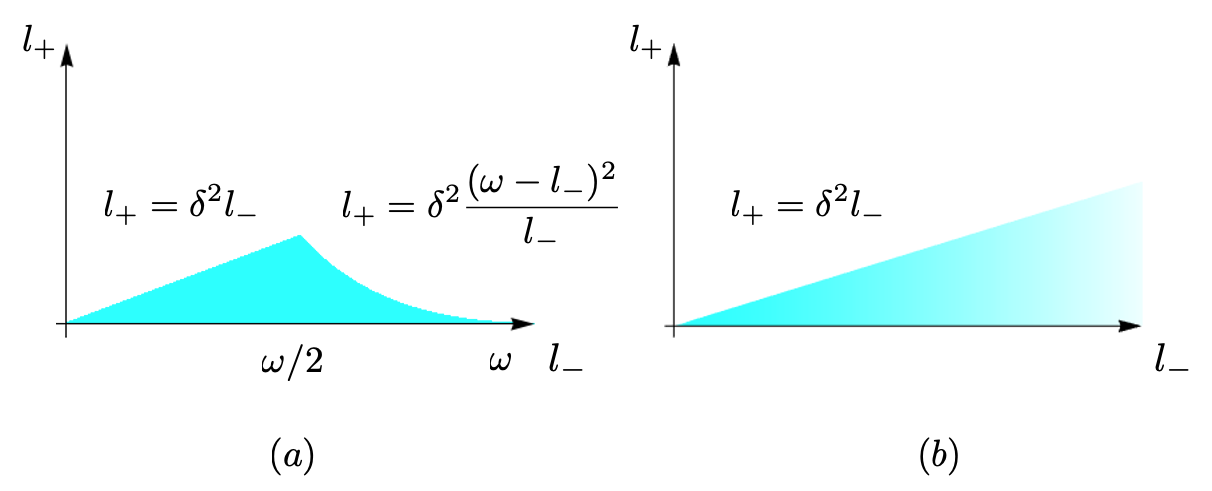}
\end{center}  
\vspace{-0.5cm}
\caption{\baselineskip 3.0ex   Phase space for the $n$-collinear jet function. (a) Naive collinear 
contribution, (b) Zero-bin contribution in the limit $l_- \rightarrow 0$. Another limit 
$l_- \rightarrow \omega$ gives the same result. \label{jetphase}}
\end{figure}

The cone jet algorithm for the $n$-collinear jet requires $\theta_1 <R$ and $\theta_2 <R$, 
where  $R$ is the jet cone size, and $\theta_i$ is the angle of the gluon $i$ with respect to 
the jet axis, which is chosen to be in the $n$ direction. This jet algorithm for the collinear part 
can be written as \cite{Chay:2015ila,Cheung:2009sg,Ellis:2010rwa}
\begin{equation} \label{cojetal}
\Theta_J =  \Theta \Bigl( \delta^2 > \frac{l_+}{l_-}\Bigr) \Theta \Bigl(l_- <\frac{\omega}{2}\Bigr) + 
\Theta \Bigl( \delta^2 > \frac{l_- l_+}{(\omega-l_-)^2}\Bigr) \Theta \Bigl(l_- >\frac{\omega}{2}\Bigr),
\end{equation}
where $\delta= \tan R/2$.  The two final-state particles should satisfy both $\theta_1 <R$ 
and $\theta_2 <R$. However, if $E_1<E_2$, that is, if $l_-<\omega/2$, when the constraint 
$\theta_1<R$ is satisfied, the condition $\theta_2<R$ is automatically satisfied, and vice versa. 
This fact is implemented in the jet algorithm, Eq.~(\ref{cojetal}). 
The above jet algorithm includes the soft modes when $l_- \rightarrow 0$ for $p_1$ or 
$l_- \rightarrow \omega$ for $p_2$.  The zero-bin subtraction should be employed to 
subtract the soft contribution from the jet function to avoid double 
counting~\cite{Manohar:2006nz}.  If we switch $p_1$ and $p_2$ in Eq.~(\ref{cojetal}), 
the two terms are switched to give the same result. And the calculation involved in the jet 
function is also invariant under this switch since the final-state particles are identical. 
Therefore the zero-bin contribution is obtained by choosing the jet algorithm for the zero-bin 
contribution from the first term in Eq.~(\ref{cojetal})  with $l_- \rightarrow 0$, and we multiply it 
by two to get the final answer. Therefore the jet algorithm for the zero-bin contribution 
is given by
\begin{equation}  \label{cozero}
\Theta_J^0 =  \Theta \Bigl( \delta^2 > \frac{l_+}{l_-}\Bigr). 
\end{equation}
The phase spaces for the naive collinear contribution and the zero-bin contribution are 
shown in Fig.~\ref{jetphase} (a), (b) respectively. Here we consider the integrated gluon 
jet function $\mathcal{J}_g$.
   
The Feynman diagrams for the gluon jet function at one loop are shown in Fig.~\ref{gluejet}. 
Figs.~\ref{gluejet} (a) and (b) are the virtual and real corrections from the Wilson lines. 
Figs.~\ref{gluejet} (c)--(e) are the cut diagrams for a fermion, a gluon and a ghost loop respectively.
The vertical dashed lines represent the unitarity cuts. We present the result by including the modified 
Wilson lines in  Eq.~(\ref{rareg}) to extract the rapidity divergence. 

The naive collinear and zero-bin contributions from Fig.~\ref{gluejet} (a) are given as
\begin{align}
\tilde{M}_a &= \frac{ig^2 C_A}{2} \Bigl( \frac{\mu^2 e^{\gamma}}{4\pi}\Bigr)^{\eps} 
\int \frac{d^D l}{(2\pi)^D} \frac{1}{l^2 (p-l)^2}
\Bigl[\frac{\omega+l_- }{\omega-l_-} \Bigl(\frac{\nu}{\omega-l_-}\Bigr)^{\eta}+\frac{2\omega-l_-}{l_-}
\Bigl(\frac{\nu}{l_-}\Bigr)^{\eta}\Bigr] \nonumber  \\
&= -\frac{\alpha_s C_A}{4\pi}   \Bigl( \frac{1}{\euv} -\frac{1}{\eir}\Bigr) 
\int_0^{\omega} dl_- \frac{2\omega-l_-}{l_-} \Bigl( \frac{\nu}{l_-}\Bigr)^{\eta},
 \\
M_a^{\varnothing}&= -2ig^2 C_A \Bigl( \frac{\mu^2 e^{\gamma}}{4\pi}\Bigr)^{\eps} 
\int \frac{d^D l}{(2\pi)^D} \frac{1}{ l^2 l_- l_+} \Bigl( \frac{\nu}{l_-}\Bigr)^{\eta}
= -\frac{\alpha_s C_A}{4\pi}  \Bigl(\frac{1}{\euv} -\frac{1}{\eir}\Bigr) \int_0^{\infty} 
dl_- \frac{2\omega}{l_-} \Bigl( \frac{\nu}{l_-}\Bigr)^{\eta}, \nonumber
\end{align}
where $\omega=p^-$, and the contour integral in the complex $l_+$-plane is performed first.
Here we use the relation
\begin{equation}
\mu^{2\eps} \int_0^{\infty} \frac{d\mathbf{l}_{\perp}^2 }{(\mathbf{l}_{\perp}^2)^{1+\eps}} 
=\frac{1}{\euv} -\frac{1}{\eir}.
\end{equation}
The net contribution is given as
\begin{align}
M_a &= \tilde{M}_a -M_a^{\varnothing} = -\frac{\alpha_s C_A}{4\pi}  \Bigl( \frac{1}{\euv} -
\frac{1}{\eir}\Bigr)   \Bigl[ \int _0^1 dx (-1) -\int_1^{\infty} dx 2x^{-1-\eta} \Bigr] \nonumber \\
&=  \frac{\alpha_s C_A}{4\pi} \Bigl(\frac{1}{\euv} -\frac{1}{\eir}\Bigr) \Bigl(\frac{2}{\eta} 
+ 2\ln \frac{\nu}{\omega} +1\Bigr),
\end{align}
where $x = l_-/\omega$.  The zero-bin contribution is divided to extract the rapidity divergence, 
and it comes from the integration of $x$ in the interval $[1,\infty]$. The spurious divergence 
at $x=0$ is cancelled when we perform the zero-bin subtraction.
 
The naive collinear contribution from Fig.~\ref{gluejet} (b) is given by
\begin{align}
\tilde{M}_b &= \frac{\alpha_s C_A}{8\pi \omega} \dms
\int  dl_- dl_+ l_+^{-1-\eps} l_-^{-\eps} \Bigl( \frac{\nu}{l_-}\Bigr)^{\eta}
\Bigl( \frac{\omega+l_-}{\omega-l_-}+\frac{2\omega-l_-}{l_-} \Bigr)  \Theta_J \nonumber \\
&=\frac{\alpha_s C_A}{4\pi} \Bigl[ \frac{1}{\eir^2} +\frac{1}{\eir}
\Bigl(1+\ln \frac{\mu^2}{\omega^2 \delta^2}\Bigr) 
+\ln \frac{\mu^2}{\omega^2 \delta^2} +\frac{1}{2}\ln^2 \frac{\mu^2}{\omega^2 \delta^2}+2
+2 \ln 2 -\frac{5}{12}\pi^2\Bigr].
\end{align}
Note that there is no rapidity divergence in the naive contribution. The zero-bin contribution, 
using Eq.~(\ref{cozero}), is given as
\begin{align} \label{jetb0}
M_b^{\varnothing} &= \frac{\alpha_s C_A}{2\pi} 
\dms \int_0^{\infty} dl_-  l_-^{-1-\eps} 
\Bigl( \frac{\nu}{l_-}\Bigr)^{\eta}\int_0^{\delta^2 l_-} dl_+  l_+^{-1-\eps}\nonumber \\
&= \frac{\alpha_s C_A}{2\pi} \frac{ e^{\gamma_{\mathrm{E}} \eps}}{\Gamma (1-\eps)} 
\Bigl(\frac{\mu^2}{\omega^2}\Bigr)^{\eps} 
\Bigl( \frac{\nu}{\omega}\Bigr)^{\eta} \int _0^{\infty} dx x^{-1-\eps-\eta} 
\int_0^{\delta^2 x} dy y^{-1-\eps},
\end{align}
where we rescale the variables as $x = l_-/\omega$, $y=l_+/\omega$.  If we integrate the above 
integral by brute force, the $y$-integral produces an IR pole $1/\eir$, and the $x$-integral reaches 
the infinity which cannot be handled by $\eir$. Therefore care must be taken 
to compute this integral.  The $x$-integral is divided into two regions with an arbitrary number 
$\kappa$ as
\begin{align} \label{bzero}
I&= \int _0^{\infty} dx x^{-1-\eps-\eta} \int_0^{\delta^2 x} dy y^{-1-\eps} = 
\Bigl[ \int_0^{\kappa} dx x^{-1-\eps-\eta} 
+\int_{\kappa}^{\infty} dx x^{-1-\eps-\eta} \Bigr] \int_0^{\delta^2 x} dy y^{-1-\eps} 
\nonumber \\
&= \int_0^{\kappa} dx x^{-1-\eps-\eta} \int_0^{\delta^2 x} dy y^{-1-\eps} 
+ \int_{\kappa}^{\infty} dx x^{-1-\eps-\eta}
\Bigl[ \int_0^{\infty} dy y^{-1-\eps} -\int_{\delta^2 x}^{\infty} dy y^{-1-\eps}\Bigr]   \\
&= \int_0^{\kappa} dx x^{-1-\eps} \int_0^{\delta^2 x} dy y^{-1-\eps} 
- \int_{\kappa}^{\infty} dx x^{-1-\eps}\int_{\delta^2 x}^{\infty} dy y^{-1-\eps}
+ \int_{\kappa}^{\infty} dx x^{-1-\eps-\eta}\int_0^{\infty} dy y^{-1-\eps}.  \nonumber
\end{align}
The first integral contains the IR divergence only, and it can be expressed in terms of the poles in $\eir$. 
Similarly, the second integral is of pure UV origin, and it can be expressed in terms of $\euv$. 
They do not contain any rapidity divergence.  The last integral involves a rapidity divergence 
because the phase space contains the region $x\rightarrow \infty$ and $y\rightarrow 0$, and it
can be evaluated by changing variables to $\alpha^2 = xy$ and $\beta^2 =y/x$. Then the third 
integral is given as
\begin{equation}
\int_0^{\infty} d\alpha^2 (\alpha^2)^{-1-\eps-\eta/2} \int_0^{\alpha/\kappa} d\beta 
\beta^{-1+\eta} = \frac{\kappa^{-\eta}}{\eta} \Bigl( \frac{1}{\euv} -\frac{1}{\eir}\Bigr).
\end{equation}
When all the terms are added in Eq.~\eqref{bzero}, the dependence on an arbitrary $\kappa$ 
cancels and the zero-bin contribution is given by
\begin{equation}
M_b^{\varnothing} = \frac{\alpha_s C_A}{4\pi} \Bigl[ -\frac{1}{\euv^2} + \frac{1}{\eir^2} +
\Bigl( \frac{1}{\euv} -\frac{1}{\eir}\Bigr) \Bigl( \frac{2}{\eta} 
+ \ln \frac{\nu^2\delta^2}{\mu^2} \Bigr),  \Bigr],
\end{equation}
and the net contribution is given as
\begin{align}
M_b = \tilde{M}_b -M_b^{\varnothing} &= \frac{\alpha_s C_A}{4\pi} 
\Bigl[ \frac{1}{\euv^2}   -\frac{1}{\euv} \Bigl(
\frac{2}{\eta} +  \ln \frac{\nu^2 \delta^2}{\mu^2}  \Bigr) 
+\frac{1}{\eir}\Bigl( 1+\frac{2}{\eta}+2\ln \frac{\nu}{\omega}\Bigr) \nonumber \\
&+ \ln \frac{\mu^2}{\omega^2 \delta^2} +\frac{1}{2} 
\ln^2 \frac{\mu^2}{\omega^2 \delta^2}+2+2 \ln 2 -\frac{5}{12}\pi^2 \Bigr].
\end{align}
If we add $M_a$ and $M_b$, we obtain the result
\begin{equation} \label{ab}
M_a + M_b = \frac{\alpha_s C_A}{4\pi} \Bigl[ \frac{1}{\euv^2} +\frac{1}{\euv} 
\Bigl(1+\ln \frac{\mu^2}{\omega^2 \delta^2}
 \Bigr) +  \ln \frac{\mu^2}{\omega^2 \delta^2} 
 +\frac{1}{2} \ln^2 \frac{\mu^2}{\omega^2 \delta^2} +2+2 \ln 2 -\frac{5}{12}\pi^2 \Bigr],
\end{equation}
which is free of the IR and the rapidity divergences.

The loops in Figs.~\ref{gluejet} (c)-(e) consist of fermions, gluons and ghost particles respectively. 
The zero-bin contributions are power suppressed compared to the naive collinear 
contributions, thus neglected. The naive collinear contributions from the fermions, 
the gluons and the ghost particles are given respectively by
\begin{align}
\tilde{M}_f &= \frac{\alpha_s T_F n_f}{2\pi \omega} \dms 
\int dl_- dl_+ l_-^{2-\eps}   \Bigl( 1-\frac{l_-}{\omega}\Bigr)^2  l_+^{-1-\eps}  
 \Bigl[ \Bigl( \frac{1}{l_-} +\frac{1}{\omega-l_-}\Bigr)^2 -\frac{2}{1-\eps} 
\frac{1}{l_- (\omega-l_-)}\Bigr] \Theta_J  \nonumber \\
&=\frac{\alpha_s T_F n_f}{4\pi} \Bigl(-\frac{4}{3\eir} -\frac{4}{3}  \ln \frac{\mu^2}{\omega^2 \delta^2}  -
\frac{23}{9} -\frac{8}{3}\ln 2\Bigr), \nonumber \\
\tilde{M}_g &=  -\frac{\alpha_s C_A}{8\pi \omega}   \dms    
\int dl_- dl_+ \Bigl[ 4 l_-^{-\eps} l_+^{-1-\eps} -\frac{5-4\eps}{1-\eps} l_-^{1-\eps} \Bigl(1-\frac{l_-}{\omega}\Bigr) \frac{l_+^{-1-\eps}}{\omega} \Bigr]\Theta_J
\nonumber \\
&=\frac{\alpha_s C_A}{8\pi} \Bigl( \frac{19}{6\eir} 
+\frac{19}{6} \ln \frac{\mu^2}{\omega^2 \delta^2}  +\frac{247}{36} +\frac{19}{3}\ln 2\Bigr), \nonumber \\
\tilde{M}_{\mathrm{ghost}} &=    -\frac{\alpha_s C_A}{8\pi \omega^2 }  \dms 
\int dl_- dl_+ l_-^{1-\eps} \Bigl(1-\frac{l_-}{\omega}\Bigr) l_+^{-1-\eps} \Theta_J\nonumber \\
&=\frac{\alpha_s C_A}{8\pi} \Bigl( \frac{1}{6\eir} +\frac{1}{6}
\ln \frac{\mu^2}{\omega^2 \delta^2} + \frac{13}{36} +\frac{1}{3}\ln 2\Bigr),
\end{align}
where $n_f$ is the number of quark flavors. The total contribution is given by
\begin{align}
\tilde{M}_f  + \tilde{M}_g +\tilde{M}_{\mathrm{ghost}}&= \frac{\alpha_s}{4\pi}
\Bigl[  \Bigl(\frac{5}{3}C_A -\frac{4}{3} T_F n_f \Bigr) \Bigl(  \frac{1}{\eir}  
+ \ln \frac{\mu^2}{\omega^2 \delta^2} \Bigr) \nonumber \\
&+\frac{65}{18} C_A -\frac{23}{9}T_F n_f +2\Bigl(\frac{5}{3}C_A -\frac{4}{3} T_F n_f \Bigr)\ln 2\Bigr].
\end{align}

Finally, the gluon field-strength renormalization at one loop is given by
\begin{equation}
Z_g^{(1)}=\frac{\alpha_s}{4\pi} \Bigl( \frac{1}{\euv} -\frac{1}{\eir}\Bigr) \Bigl(\frac{5}{3}C_A 
-\frac{4}{3} T_F n_f \Bigr) . 
\end{equation}
The overall contribution of the real and virtual corrections from the gluon self energy at order 
$\alpha_s$ is given by
\begin{align}
 M_{\mathrm{self}} &=\frac{\alpha_s}{4\pi} \Bigl[ \Bigl(\frac{5}{3}C_A -\frac{4}{3} T_F n_f \Bigr) 
 \Bigl(  \frac{1}{\euv}  + \ln \frac{\mu^2}{\omega^2 \delta^2} \Bigr)   \nonumber \\
&+\frac{65}{18} C_A -\frac{23}{9}T_F n_f +2\Bigl(\frac{5}{3}C_A -\frac{4}{3} T_F n_f \Bigr) \ln 2\Bigr].
\end{align}

The total collinear contributions are given by
\begin{align}
M_{\mathrm{coll}} &= 2(M_a +M_b) +M_{\mathrm{self}}  \nonumber \\
&= \frac{\alpha_s}{2\pi} \Bigl[ \frac{C_A}{\euv^2} +\frac{1}{\euv} \Bigl( \frac{\beta_0}{2} +C_A
\ln \frac{\mu^2}{\omega^2 \delta^2}  \Bigr) +\frac{\beta_0}{2}\ln \frac{\mu^2}{\omega^2 \delta^2}
+\frac{1}{2} C_A \ln^2 \frac{\mu^2}{\omega^2 \delta^2} \nonumber \\
&+ \frac{137}{36} C_A -\frac{23}{18} T_F n_f +\beta_0 \ln 2 -\frac{5}{12}C_A \pi^2\Bigr],
\end{align}
where $\beta_0 =11C_A/3- 4 T_F n_f/3$ is the leading term of the QCD beta function. [See Eq.~\eqref{betaq}.]
The collinear contribution is clearly IR finite, and 
the renormalized gluon jet function at one loop is obtained by adding the counterterms as
\begin{equation}
\mathcal{J}_g^{(1)} (\omega \delta, \mu) =  \frac{\alpha_s}{4\pi} \Bigl( \beta_0
\ln \frac{\mu^2}{\omega^2 \delta^2}+  C_A \ln^2 \frac{\mu^2}{\omega^2 \delta^2}  + \frac{137}{18} C_A 
-\frac{23}{9} T_F n_f + 2\beta_0 \ln 2
-\frac{5}{6}C_A \pi^2\Bigr).
\end{equation}
This coincides with the result on the unmeasured gluon jet function obtained in Ref.~\cite{Ellis:2010rwa}.
 And the anomalous dimension of the gluon jet function at NLO is given by
\begin{equation} \label{anomjet}
\gamma_{J} = \frac{d}{d\ln \mu} \mathcal{J}_g = \frac{\alpha_s}{\pi} \Bigl( C_A \ln \frac{\mu^2}{\omega^2 \delta^2} +\frac{\beta_0}{2}\Bigr).
\end{equation}

\subsection{Quark beam function\label{cdfn}}
 
\begin{figure}[t] 
\begin{center}
\includegraphics[height=3.5cm]{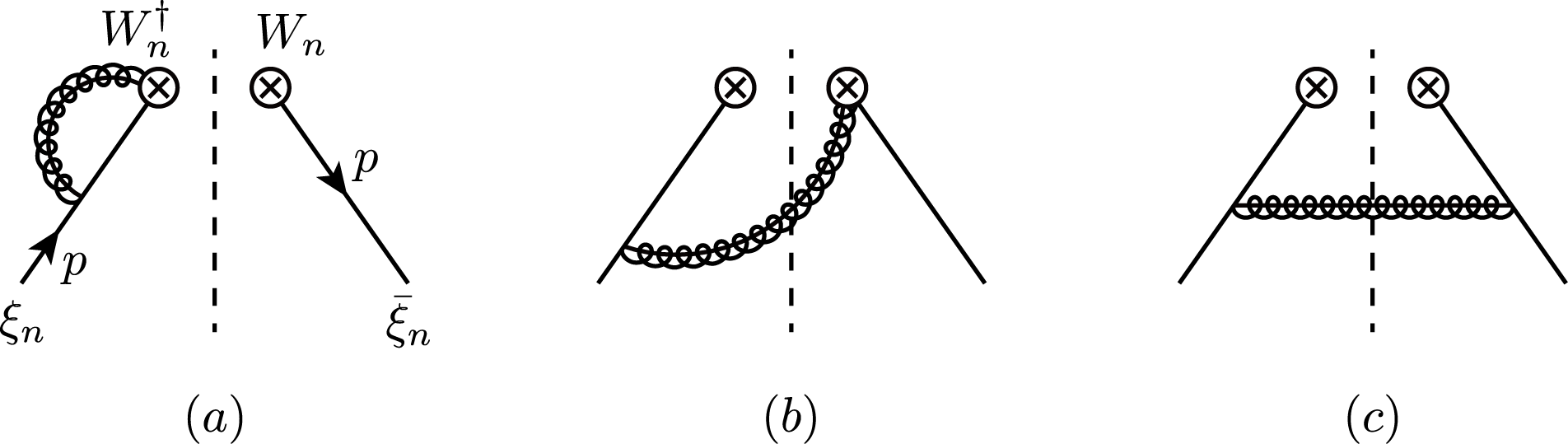}
\end{center}  
\vspace{-0.3cm}
\caption{\baselineskip 3.0ex Feynman diagrams for the $n$-collinear beam function  at one loop  
(a) virtual corrections, (b) and (c) real gluon emission. The mirror images of (a)
and (b) are omitted. the dashed lines denote the final-state cuts.\label{cof}}
\end{figure}

The beam jets are produced in the beam directions from the initial-state radiation, and 
the evolution of the initial-state particles from $Q\lambda$ to $\mu$ is described by 
the beam function.  The quark beam function and the unmeasured quark beam function are defined in Eqs.~\eqref{beamf} and 
\eqref{unbeam}, and we compute the unmeasured quark beam function here. 
 The Feynman diagrams for the beam functions
at one loop are shown in Fig.~\ref{cof}.  Fig.~\ref{cof} (a) is the virtual correction, 
Fig.~\ref{cof} (b) and (c) are real gluon emissions.  

 The naive collinear contribution from Fig.~\ref{cof} (a)  is given as
\begin{align} \label{ncdf}
\tilde{M}_a &= 2ig^2 C_F \Bigl( \frac{\mu^2 e^{\gamma}}{4\pi}\Bigr)^{\eps} \delta (1-z)  \int \frac{d^D l}{(2\pi)^D} 
\frac{\omega -l_-}{l^2 (l-p)^2  l_-}  \Bigl(\frac{\nu}{l_-} \Bigr)^{\eta}\nonumber \\
&=-\frac{\alpha_s C_F}{2\pi} \Bigl(\frac{\nu}{\omega} \Bigr)^{\eta}\delta (1-z) 
\Bigl( \frac{1}{\euv} -\frac{1}{\eir}\Bigr) \int_0^1 dx \frac{1-x}{x^{1+\eta}}, 
\end{align}
where $x= l_-/\omega$. There is unwanted divergence as $x\rightarrow 0$, but it is cancelled by the zero-bin 
contribution which is given as
\begin{equation}
M_a^{\varnothing} = -\frac{\alpha_s C_F}{2\pi} \Bigl(\frac{\nu}{\omega} \Bigr)^{\eta}\delta (1-z) 
\Bigl( \frac{1}{\euv} -\frac{1}{\eir}\Bigr) \int_0^{\infty} dx \frac{1}{x^{1+\eta}}.
\end{equation}
As in the calculation of the beam function, the net collinear contribution from Fig.~\ref{cof} (a) is given by
\begin{equation}
M_a = \tilde{M}_a -M_a^{\varnothing} = \frac{\alpha_s C_F}{2\pi} \delta (1-z) 
\Bigl( \frac{1}{\euv} -\frac{1}{\eir}\Bigr) \Bigl( \frac{1}{\eta} + \ln \frac{\nu}{\omega} +1\Bigr).
\end{equation}

The naive contribution from Fig.~\ref{cof} (b) is written as
\begin{align}
\tilde{M}_b &= -4\pi g^2 C_F \Bigl( \frac{\mu^2 e^{\gamma}}{4\pi}\Bigr)^{\eps} 
\int dt \int \frac{d^D l}{(2\pi)^D}
\frac{\omega- l_-}{(l-p)^2 l_-} \Bigl(\frac{\nu}{l_-} \Bigr)^{\eta}\delta (l^2) \delta 
\Bigl( 1-z-\frac{l_-}{\omega}\Bigr) \delta (t - \omega l_+) \Theta_B\nonumber \\
&= \frac{\alpha_s C_F}{2\pi}  \Bigl(\frac{\nu}{\omega} \Bigr)^{\eta}\dms \int_0^{z (1-z) 
\delta^2 \omega^2}  dt\,  t^{-1-\eps}  
z^{1+ \eps} (1-z)^{-1- \eps-\eta}.
\end{align}
We treat $\eta$ much smaller than $\eps$, hence $\eta$ can be neglected. Then $\tilde{M}_b$ can be 
written as
\begin{equation}
\tilde{M}_b = -\frac{\alpha_s C_F}{2\pi} \dms \frac{1}{\eir} (\delta^2 \omega^2)^{-\eps} z (1-z)^{-1-2\eps}.
\end{equation}
Here we use the relation
\begin{equation}
(1-z)^{-1-2 \eps} =  -\frac{1}{2\eir} \delta (1-z) +\mathcal{L}_0 (1-z) -2 \eir \mathcal{L}_1 (1-z) +\cdots,
\end{equation}
where the functions $\mathcal{L}_n (z)$ are defined as
\begin{equation}\label{ldef}
\mathcal{L}_n (z) \equiv \Bigl[ \frac{\theta (z) \ln^n z}{z}\Bigr]_+ = \lim_{\beta\rightarrow 0} 
\Bigl[ \frac{\theta (z-\beta) \ln^n z}{z} +\delta (z-\beta) \frac{\ln^{n+1} \beta}{n+1}\Bigr].
\end{equation} 
Then the naive contribution contains only the IR divergence and is given by
\begin{align}
\tilde{M}_b &=\frac{\alpha_s C_F}{2\pi} \Bigl[ \Bigl( \frac{1}{2\eir^2} +
 \frac{1}{2\eir} \ln \frac{\mu^2}{\omega^2 \delta^2} 
+\frac{1}{4} \ln^2 \frac{\mu^2}{\omega^2 \delta^2} -\frac{\pi^2}{24}\Bigr) \delta (1-z) 
 \nonumber \\
&-\frac{1}{\eir} z \mathcal{L}_0 (1-z)-z \mathcal{L}_0 (1-z) \ln \frac{\mu^2}{\omega^2 \delta^2}   
+ 2z \mathcal{L}_1 (1-z)\Bigr].
\end{align}

The zero-bin contribution $M_b^{\varnothing}$ is written as
\begin{align} \label{beamb0}
 M_{b}^{\varnothing} &= \frac{\alpha_s C_F}{2\pi} \dms \int dl_+ dl_- \frac{1}{(l_+ l_-)^{1+\eps}} 
 \Bigl(\frac{\nu}{l_-} \Bigr)^{\eta} \Theta_B \nonumber \\
 &= \frac{\alpha_s C_F}{2\pi} \frac{e^{\gamma \eps}}{\Gamma(1-\eps)} 
 \Bigl(\frac{\mu^2}{\omega^2}\Bigr)^{\eps}
 \Bigl(\frac{\nu}{\omega}\Bigr)^{\eta} \int_0^{\infty} dx \, x^{-1-\eps-\eta} 
 \int_0^{\delta^2 x} dy \, y^{-1-\eps}
 \nonumber \\
 &= \frac{\alpha_s C_F}{2\pi} \delta (1-z) \Bigl[ -\frac{1}{2} \Bigl( \frac{1}{\euv^2} 
 -\frac{1}{\eir^2}\Bigr) 
 +\frac{1}{2} \Bigl( \frac{1}{\euv} -\frac{1}{\eir}\Bigr) \Bigl( \frac{2}{\eta} 
 +\ln \frac{\nu^2 \delta^2}{\mu^2} \Bigr)
 \Bigr].
 \end{align}
 The net contribution is given by
\begin{align}
M_b &= \tilde{M}_b -M_b^{\varnothing} \nonumber \\
&= \frac{\alpha_s C_F}{2\pi} \Bigl\{ \delta (1-z)\Bigl[ \frac{1}{2\euv^2} 
+\frac{1}{2\euv} \ln \frac{\mu^2}{\omega^2 \delta^2} + \frac{1}{4} \ln^2 
\frac{\mu^2}{\omega^2 \delta^2}
-\Bigl( \frac{1}{\euv} -\frac{1}{\eir}\Bigr)  \Bigl( \frac{1}{\eta} 
+ \ln \frac{\nu}{\omega} -\frac{\pi^2}{24}\Bigr)
\Bigr] \nonumber \\
&-\Bigl( \frac{1}{\eir} + \ln \frac{\mu^2}{\omega^2 \delta^2}\Bigr) z \mathcal{L}_0 (1-z)  
+ 2z \mathcal{L}_1 (1-z) \Bigr\}.
\end{align}

The contribution of Fig.~\ref{cof} (c) is given as
\begin{align}
M_c &= 4\pi g^2 C_F (1-\eps)  \ms \int dt \int \dd{l} \frac{\mathbf{l}_{\perp}^2}{[(l-p)^2]^2} 
\delta (l^2) \delta (t-\omega l_+ )  \omega \delta (\omega - p_- + l_-)\nonumber \\
&=-\frac{\alpha_s C_F}{2\pi} (1-z) \Bigl( \frac{1}{\eir} -1 
+\ln \frac{\mu^2}{\omega^2 \delta^2} -2 \ln (1-z)\Bigr).  
\end{align}
The zero-bin contribution is suppressed and is neglected.

 By including the wave function renormalization, the beam function at NLO is given by
 \begin{align}
& B^{(1)} (z, \delta, \mu) = 2(M_a + M_b) + M_c -\frac{\alpha_s C_F}{4\pi} \delta (1-z) \Bigl( \frac{1}{\euv}
 -\frac{1}{\eir} \Bigr)  \\
 &= \frac{\alpha_s C_F}{2\pi} \Bigl\{ \delta (1-z) \Bigl[ \frac{1}{\euv^2} +\frac{1}{\euv} \Bigl(
 \ln \frac{\mu^2}{\omega^2 \delta^2} +\frac{3}{2}\Bigr) +\frac{1}{2} \ln^2 \frac{\mu^2}{\omega^2 \delta^2} 
 +\frac{3}{2} \ln \frac{\mu^2}{\omega^2 \delta^2}  -\frac{\pi^2}{12}   \Bigr] \nonumber \\
&  -P_{qq} (z) \Bigl( \frac{1}{\eir} +\ln \frac{\mu^2}{\omega^2 \delta^2} \Bigr)
 + 4z \mathcal{L}_1 (1-z) + (1-z) \Bigl( 1 +2 \ln (1-z)\Bigr)  \Bigr\}, \nonumber
  \end{align}
where $P_{qq} (z)$ is the splitting function, which is defined as
\begin{equation}
P_{qq}(z) = \frac{3}{2} \delta (1-z) + 2z \mathcal{L}_0 (1-z) + (1-z).
\end{equation}
Note that the beam function contains the IR divergence, which is the same as the PDF, and 
it is absorbed into the nonperturbative matrix element
in the beam function in exactly the same way as we treat the PDF.   

The renormalized beam function at NLO by removing the divergent terms is given by
\begin{equation}
  B_R^{(1)} (z, \omega\delta, \mu)  
= \frac{\alpha_s C_F}{2\pi} \Bigl( \frac{1}{2} \ln^2 \frac{\mu^2}{\omega^2 \delta^2} 
 +\frac{3}{2} \ln \frac{\mu^2}{\omega^2 \delta^2} +8-\frac{3}{4}\pi^2\Bigr),
 \end{equation}
 and the anomalous dimension of the beam function at NLO is given as
 \begin{equation}
 \gamma_{B}^{(1)} = \mu \frac{d}{d\mu} B^{(1)} =\frac{\alpha_s}{\pi} 
 \ln \frac{\mu^2}{\omega^2 \delta^2} + \frac{\alpha_s}{\pi} \frac{3}{2} C_F.
 \end{equation}
 
\section{Soft function\label{sofun}}
The soft function, defined in Eq.~(\ref{softfunction}), is given again by
\begin{equation} 
S_{JI} = \sum_{X_s}  \mathrm{tr} \langle 0| Y_1^{\dagger} (\mathcal{Y}_3^{\dagger} T_J^{\dagger} 
\mathcal{Y}_4)^{ba} Y_2 |X_s
\rangle \Theta_J \Theta_B\langle X_s |Y_2^{\dagger} (\mathcal{Y}_4^{\dagger} T_I 
\mathcal{Y}_3)^{ab} Y_1|0\rangle.
\end{equation}
The Wilson lines $Y_2^{\dagger}$ and $Y_1$ correctly describe the soft gluon emission from 
the antiquark and the quark in the initial state. But, to be exact, 
the Wilson lines $\mathcal{Y}_4^{\dagger}$ and $\mathcal{Y}_3$ should be 
$\tilde{\mathcal{Y}}_4^{\dagger}$ and $\tilde{\mathcal{Y}}_3$,
describing the soft gluon emission from the outgoing particles \cite{Chay:2004zn}. However, 
we keep the forms as they are for notational simplicity. The soft Wilson lines are given as
\begin{align}
&Y_1  = \sum_{\mathrm{perm.}} \exp \Bigl[ \frac{1}{n_1\cdot \mathcal{R} + i0} 
(-g n_1 \cdot A_s)\Bigr],  \ Y_2^{\dagger} = 
\sum_{\mathrm{perm.}} \exp \Bigl[ -g n_2 \cdot A_s \frac{1}{n_2 \cdot \mathcal{R}^{\dagger} -i0}\Bigr],  
\nonumber \\
& \mathcal{Y}_3 =  \sum_{\mathrm{perm.}} \exp \Bigl[ \frac{1}{n_3\cdot \mathcal{R} -i0} 
(-gn_3\cdot A_s)\Bigr],  \ 
\mathcal{Y}_4^{\dagger} =  \sum_{\mathrm{perm.}} \exp \Bigl[ -g n_4 
\cdot A_s \frac{1}{n_4 \cdot \mathcal{R}^{\dagger} +i0}\Bigr],
\end{align}
where $\mathcal{R}$ is the operator extracting the soft momentum. The rapidity regulator cannot be expressed 
in a general form, and it will be presented in the explicit calculation order by order.
 
\begin{figure}[b] 
\begin{center}
\includegraphics[width=8cm]{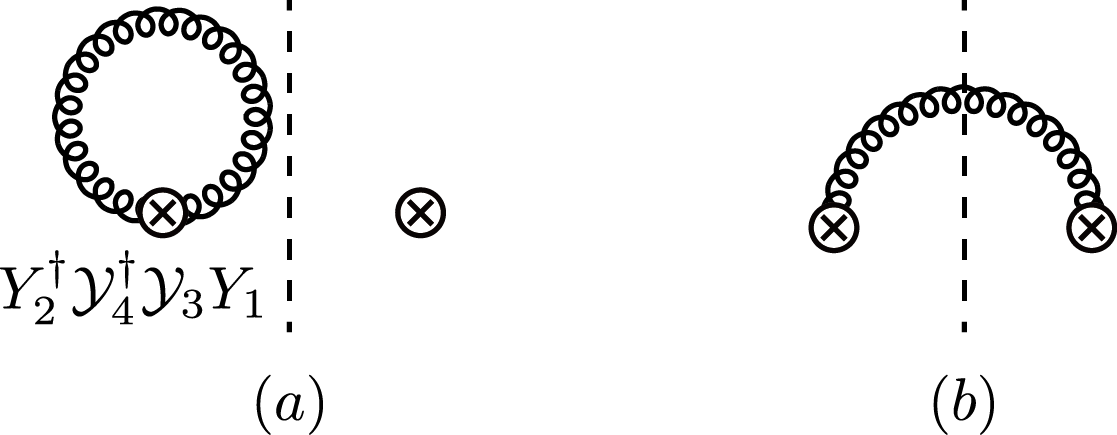}
\end{center}  
\vspace{-0.5cm}
\caption{\baselineskip 3.0ex   Feynman diagrams for the soft function at one loop. 
(a) Virtual correction (b) Real gluon emission. The hermitian conjugates are omitted. 
The dashed lines are the unitarity cuts.
\label{feynmansoft}}
\end{figure}

The Feynman diagrams for the soft function at NLO is shown in Fig.~\ref{feynmansoft}. 
Figs.~\ref{feynmansoft} (a) and (b) represent the virtual and the real corrections respectively. 
The vertical dashed lines are the unitarity cuts, and the hermitian conjugates are 
not shown.  Fig.~\ref{feynmansoft} (a) yields
\begin{equation} \label{softvirt}
M_{ij}^V = -2\pi g^2 \ms \int \dd{l} \frac{2n_{ij}}{l_i l_j} \delta (l^2) R(l_i, l_j), 
\end{equation}
apart from the group theory factors 
$-\mathbf{T}_i \cdot \mathbf{T}_j$~\cite{Chiu:2009mg,Catani:1996jh,Catani:1996vz}. 
Here $n_{ij} = n_i \cdot n_j/2$, $l_i = n_i \cdot l$ and $l_j = n_j \cdot l$. $R(l_i, l_j)$ is 
the rapidity regulator, which can be written at order $\alpha_s$ as
\begin{equation} \label{rapred}
R(l_i, l_j) = \Bigl( \frac{\nu n_{ij}}{l_i}\Bigr)^{\eta} \theta (l_i-l_j)+
\Bigl( \frac{\nu n_{ij}}{l_j}\Bigr)^{\eta} \theta (l_j-l_i).
\end{equation} 
 In the basis of $n_i$ and $n_j$, the momentum $l_i^{\mu}$ can be written as
 \begin{equation}
 l^{\mu} = l_j \frac{n_i^{\mu}}{2n_{ij}} + l_i \frac{n_j^{\mu}}{2n_{ij}} +l_{ij,\perp}^{\mu},
 \end{equation}
 where $l_{ij,\perp}^{\mu}$ is the momentum perpendicular to $n_i$ and $n_j$, with 
 $l^2 = l_i l_j/n_{ij} + l_{ij,\perp}^2$.
  
 Since Eq.~\eqref{softvirt} is symmetric under the exchange $i\leftrightarrow j$, we pick up the first part of the
 rapidity regulator in Eq.~\eqref{rapred} and multiply two. Then the virtual contribution is given as
\begin{align}
M_{ij}^V &= -\frac{\alpha_s}{\pi} \dms \int dl_i dl_j \Bigl( \frac{l_i l_j}{n_{ij}} \Bigr)^{-\eps} \frac{1}{l_i l_j}
\Bigl(\frac{n_{ij} \nu}{l_i} \Bigr)^{\eta} \theta (l_i > l_j) \nonumber \\
&=-\frac{\alpha_s}{\pi} \dms n_{ij}^{\eps+\eta} \nu^{\eta} \int_0^{\infty} dl_i l_i^{-1-\eps-\eta}
\int_0^{l_i} dl_j l_j^{-1-\eps}.
\end{align}
In order to extract the divergences correctly, we separate the integration regions by introducing an arbitrary
intermediate scale $\Omega$ as
\begin{align}
I &= \int_0^{\infty} dl_i l_i^{-1-\eps-\eta} \int_0^{l_i} dl_j l_j^{-1-\eps} \nonumber \\
&= \int_0^{\Omega} dl_i l_i^{-1-\eps-\eta} \int_0^{l_i} dl_j l_j^{-1-\eps} 
+\int_{\Omega}^{\infty} dl_i l_i^{-1-\eps-\eta} \Big[\int_0^{\infty} dl_j l_j^{-1-\eps} -\int_{l_i}^{\infty} 
dl_j l_j^{-1-\eps}\Bigr].
\end{align}
The first (third) term yields the IR (UV) divergence, and the second term is computed by 
changing variables $l_i l_j = \alpha^2$ and $l_j/l_i =\beta^2$ to extract the rapidity 
divergence. The overall result is independent of $\Omega$, and the result is given by 
\begin{equation}
M_{ij}^V= \frac{\alpha_s}{2\pi} \Bigl[ \frac{1}{\euv^2} -\frac{1}{\eir^2} 
+\Bigl( \frac{1}{\euv} -\frac{1}{\eir}\Bigr) \Bigl(-\frac{2}{\eta} 
+\ln \frac{\mu^2}{\nu^2 n_{ij}}\Bigr) \Bigr].
\end{equation}

The jet algorithm and the veto should be applied to the real gluon emission. The main 
contribution comes from the configuration in which the gluon is emitted from the soft 
Wilson lines with the eikonal factor $1/n_i \cdot l$ or $1/n_j \cdot l$ and the jet algorithm 
or the beam veto is applied with respect to the $n_i$ or $n_j$ direction. We will denote these 
contributions as $M_{ij,i}^R$ and $M_{ij,j}^R$ respectively. The contribution $M_{ij,k}^R$,
in which the $n_k$ direction is neither $n_i$ nor $n_j$, is not correlated to the $n_k$ direction, 
hence it is proportional to the area of the jet cone, $\delta^2$, and is neglected.  

The real contribution from the jet algorithm is given as
\begin{equation}
M_{ij,i}^R = 2\pi g^2 \ms \int \dd{l} \frac{2n_{ij}}{n_i\cdot l  n_j \cdot l} \delta (l^2)  
\Bigl( \frac{\nu}{\overline{n}_i \cdot l}\Bigr)^{\eta} \Theta( n_i \cdot l < 
\delta_i^2 \overline{n}_i \cdot l) \Theta (l^0 > \Lambda),
\end{equation}
where the rapidity regulator is written as $(\nu/\overline{n}_i\cdot l)^{\eta}$ because the soft gluon
is in the collinear region due to the jet algorithm. For the same reason, $n_j \cdot l$ can be expressed as
\begin{equation}
n_j \cdot l = n_j\cdot \Bigl( \overline{n}_i \cdot l \frac{n_i}{2} + l_{i\perp} 
+ n_i \cdot l \frac{\overline{n}_i}{2}\Bigr) \approx n_{ij} \overline{n}_i\cdot l. 
\end{equation}
Here the size of the jet cone in the $n_i$ direction is denoted as $\delta_i$ to distinguish 
the jet radius ($i=3, 4$) and the beam veto ($i=1, 2$) with $\delta_{1,2} = e^{-y_{\mathrm{cut}}}$. 
Then, by writing $\overline{n}_i \cdot l = l_-$ and $n_i \cdot l = l_+$,  $M_{ij,i}^R$ is written as
\begin{align}
M_{ij,i}^R &= \frac{\alpha_s}{2\pi} \dms \nu^{\eta} \int dl_- dl_+ l_-^{-1-\eps-\eta} l_+^{-1-\eps} 
\Theta (l_+ < l_- \delta_i^2) \Theta(l_-  > 2 \Lambda) \nonumber \\
&= \frac{\alpha_s}{2\pi} \dms \nu^{\eta} \int_{2\Lambda}^{\infty} dl_-  l_-^{-1-\eps-\eta}
\int_0^{\delta_i^2 l_-} dl_+  l_+^{-1-\eps} \nonumber \\
&= \frac{\alpha_s}{2\pi} \Bigl[ \Bigl( \frac{1}{\euv} -\frac{1}{\eir}\Bigr) 
\Bigl( \frac{1}{\eta} +\ln \frac{\nu}{2\Lambda}\Bigr)-\frac{1}{2\euv^2} -\frac{1}{\euv} 
\ln \frac{\mu}{2\Lambda \delta_i} -\ln^2 \frac{\mu}{2\Lambda \delta_i} \Bigr].
\end{align}

The contribution of $M_{ij,k}^R$ can be written as
\begin{equation}
M_{ij,k}^R = 2\pi g^2 \ms \int \dd{l} \frac{2n_{ij}}{n_i\cdot l n_j\cdot l} \delta (l^2)
\Theta (n_k \cdot l < \delta_k^2 
\overline{n}_k\cdot l) \Theta(\overline{n}_k \cdot l > 2\Lambda). 
\end{equation}
Because there is no rapidity divergence in the integral, the rapidity regulator is absent. Since 
the gluon momentum is $n_k$-collinear, $n_i \cdot l$ and $n_j\cdot l$ can be written as
\begin{equation}
n_i \cdot l \approx n_i \cdot \frac{n_k}{2} \overline{n}_k \cdot l = n_{ik} l_-, \ \ n_j \cdot l \approx 
n_{jk} l_-,
\end{equation}
where $l_- = \overline{n}_k \cdot l$. Then $M_{ij,k}^R$ is given as
\begin{align}
M_{ij,k}^R &= \frac{\alpha_s}{2\pi} \frac{n_{ij}}{n_{ik} n_{jk}} \dms \int dl_- dl_+ 
\frac{(l_- l_+)^{-\eps}}{l_-^2} \Theta(l_+ < \delta_k^2 l_-)  \Theta (l_- > 2\Lambda) \nonumber \\
&= \frac{\alpha_s}{2\pi} \frac{n_{ij}}{n_{ik} n_{jk}} \dms \int_{2\Lambda}^{\infty} dl_-  (l_-)^{-2-\eps}
\int_0^{\delta_k^2 l_-} dl_+ (l_+)^{-\eps}\nonumber \\
&= \frac{\alpha_s}{4\pi} \frac{n_{ij}}{n_{ik} n_{jk}} \delta_k^2 \Bigl( \frac{1}{\euv} +1 
+ 2 \ln \frac{\mu}{2\Lambda} -\ln \delta_k^2\Bigr).
\end{align}
As claimed before, it is proportional to $\delta_k^2$ and is neglected.

The soft part $M_{ij}^{\mathrm{veto}}$ is given by
\begin{equation} \label{softcal}
M_{ij}^{\mathrm{veto}} = g^2 \Bigl(\frac{\mu^2 e^{\gamma}}{4\pi}\Bigr)^{\eps} 
\int \frac{d^D l}{(2\pi)^D} \frac{n_i \cdot n_j}{n_i \cdot l n_j \cdot l} 2\pi \delta (l^2) \Theta 
(l_0 < \Lambda).
\end{equation}
The integration can be performed by choosing the lightcone vectors as
\begin{equation}
n_i = (1, 0, 0, 1), \ \ n_j = (1, \sin \theta, 0, \cos \theta),
\end{equation}
We can express the momentum $l^{\mu}$ in the $r_i$ basis as
\begin{equation}
 l^{\mu} = l_0 r_0^{\mu} + l_1 r_1^{\mu} + l_3 r_3^{\mu} + l_{\perp}^{\mu}, 
\end{equation}
where $r_0^{\mu} = (1, 0, \mathbf{0},0)$, $r_1^{\mu} = (0, 1, \mathbf{0}, 0)$, 
$r_3^{\mu} = (0, 0, \mathbf{0}, 1)$, and $l_{\perp}^{\mu}$ is perpendicular to $r_i$. The point of choosing these
bases is that $n_i\cdot l$ and $n_j\cdot l$ have nonzero components in the 0, 1 and 3 directions. And all the other
components reside in the $1-2\eps$ dimensions, which we call the perpendicular direction. The integration measure 
$d^D l$ can be written as
\begin{equation}
d^D l = dl_0 dl_1 dl_3 d^{1-2\eps} \mathbf{l}_{\perp} = dl_0 dl_1 dl_3 d\mathbf{l}^2_{\perp}
(\mathbf{l}^2_{\perp})^{1/2-\eps} \frac{d\Omega_{1-2\eps}}{2} = \frac{\pi^{1/2-\eps}}{\Gamma(\frac{1}{2}-\eps)}
dl_0 dl_1 dl_3 d\mathbf{l}^2_{\perp} (\mathbf{l}^2_{\perp})^{1/2-\eps},
\end{equation}
where the angular integral over $\Omega_{1-2\eps}$ is performed.
 
The soft part $M_{ij}^{\mathrm{veto}}$ is written as
\begin{align}
M_{ij}^{\mathrm{veto}} &= \frac{\alpha_s}{2\pi} 
\frac{(e^{\gamma_{\mathrm{E}}} \mu^2)^{\eps}}{\sqrt{\pi} \Gamma(\frac{1}{2}-\eps)}
\int dl_0 dl_1 dl_3 d\mathbf{l}^2_{\perp} (\mathbf{l}^2_{\perp})^{1/2-\eps}
\frac{1-\cos\theta}{(l_0 -l_3)(l_0 -l_3\cos\theta -l_1 \sin \theta)} \nonumber \\
&\times \delta( l_0^2 -l_1^2 -l_3^2 -\mathbf{l}_{\perp}^2) \theta (l_0) \theta (\Lambda-l_0) \nonumber \\
&=\frac{\alpha_s}{2\pi} 
\frac{(e^{\gamma_{\mathrm{E}}} \mu^2)^{\eps}}{\sqrt{\pi} \Gamma(\frac{1}{2}-\eps)}
\int dl_0 dl_1 dl_3 \frac{(1-\cos \theta)
(l_0^2 -l_1^2 -l_3^2)^{-1/2-\eps}}{(l_0 -l_3)(l_0 -l_3\cos\theta -l_1 \sin \theta)}\theta (l_0) \theta (\Lambda-l_0)
\end{align}
Now we change the variables $l_3 = l_0 k \cos \phi$ and $l_1 = l_0 \sin \phi$ with 
$dl_0 dl_3 dl_1 = dl_0 dk d\phi k l_0^2$ to write $M_{ij}^{\mathrm{veto}}$ as
\begin{align}
M_{ij}^{\mathrm{veto}} &= \frac{\alpha_s}{2\pi} 
\frac{(e^{\gamma_{\mathrm{E}}} \mu^2)^{\eps}}{\sqrt{\pi} \Gamma(\frac{1}{2}-\eps)}
\int_0^{\Lambda} dl_0 l_0^{-1-2\eps} \int_0^1 dk k (1-k^2)^{-1/2-\eps} \int_0^{2\pi}
\frac{d\phi}{(1-k\cos \phi)(1-\cos (\phi-\theta))} \nonumber \\
&= \frac{\alpha_s (1-\cos\theta)}{\sqrt{\pi}} 
\frac{(e^{\gamma_{\mathrm{E}}} \mu^2)^{\eps}}{\sqrt{\pi} \Gamma(\frac{1}{2}-\eps)} \int_0^{\Lambda} 
dl_0 l_0^{-1-2\eps} \int_0^1 dk \frac{k(1-k^2)^{-1-\eps}}{1-k^2 (1+\cos\theta)/2} \nonumber \\
&= \frac{\alpha_s}{2\pi} \Bigl( \frac{1}{\eir^2} +\frac{1}{\eir} \ln \frac{\mu^2}{4n_{ij} \Lambda^2}
+\frac{1}{2}  \ln^2 \frac{\mu^2}{4n_{ij} \Lambda^2} +Li_2 (1-n_{ij}) -\frac{\pi^2}{4}\Bigr).
\end{align}

The total soft contribution at next-to-leading order is given by
\begin{align} \label{totalsoft}
S_{ij}&= M_{ij}^V+ M_{ij,i}^R + M_{ij,j}^R + M_{ij}^{\mathrm{veto}} \nonumber \\
&= \frac{\alpha_s}{2\pi} \Bigl( \frac{1}{\euv} \ln \frac{\delta_i \delta_j}{n_{ij}} 
+ 2\ln \frac{\mu}{2\Lambda} \ln \frac{\delta_i \delta_j}{n_{ij}} 
-\ln^2 \delta_i -\ln^2 \delta_j +Li_2(1-n_{ij}) -\frac{\pi^2}{4}\Bigr).
\end{align}
The soft function is given by $-\sum_{i,j} \mathbf{T}_i \cdot \mathbf{T}_j S_{ij}$, and 
it can be written as a matrix in the operator basis.

 \section{Evolution of the scattering cross section\label{evol}}
The dijet cross section is written as
\begin{align} \label{nlsig}
\sigma &= \frac{1}{64 \pi N_c^2 S^2} \sum_{IJ}  \int dt H_{IJ} (\mu) S_{JI} (\mu)  
\int \frac{dz_1}{z_1} B_{q/N_1} (z_1, E_{\mathrm{cm}} e^{-y_{cut}}, \mu) \nonumber \\
&\times  
\int \frac{dz_2}{z_2} B_{\bar{q}/N_2} (z_2, E_{\mathrm{cm}} e^{-y_{cut}},\mu) 
\mathcal{J}_{g_3} (\omega_3 \delta_3, \mu)   \mathcal{J}_{g_4} 
(\omega_4 \delta_4, \mu).
\end{align}
The cross section is the product of the hard function $H_{IJ}$, the soft function $S_{JI}$, 
the jet functions and the beam functions. Each function starts from the characteristic scales 
$\mu_i$, at which the logarithms become small, and scales to the common factorization 
scale $\mu_F$. The characteristic scales are given as $\mu_H \sim \omega$ for the hard part, 
$\mu_J \sim \omega \delta $ and $\mu_B \sim \omega \delta$ for the collinear parts, 
and $\mu_S \sim \Lambda$ for the soft part. The overall cross section is independent of 
the factorization scale $\mu_F$. The evolution of each component in the jet cross section 
can be obtained by solving the RG equations.

\subsection{The hard function}
The hard function is given by $H_{IJ} = C_I C_J^*$, where $C_I$ is the Wilson coefficient of $O_I$ 
obtained by the matching between the full QCD and SCET. The detailed form of the hard 
function at one loop for $2\rightarrow 2$ partonic processes can be found in Ref.~\cite{Kelley:2010fn}.
The relation between the bare and the renormalized hard functions is given by
\begin{equation}
\mathbf{H}_{\mathrm{bare}} = \mathbf{Z}_H (\mu) \mathbf{H} (\mu) 
\mathbf{Z}_H^{\dagger} (\mu), 
\end{equation}
and the renormalization group equation for the hard function is given by
\begin{equation} \label{hardrg}
\frac{d}{d\ln \mu} \mathbf{H} = \mathbf{\Gamma}_H \mathbf{H} 
+ \mathbf{H}\mathbf{\Gamma}_H ^{\dagger},
\end{equation}
where the anomalous dimension matrix $\mathbf{\Gamma}_H$ is defined as
\begin{equation}
\mathbf{\Gamma}_H = -\mathbf{Z}_H^{-1} \frac{d}{d\ln \mu} \mathbf{Z}_H.
\end{equation}

The anomalous dimension matrix $\mathbf{\Gamma}_H$ can be written 
as~\cite{Kelley:2010fn,Hornig:2016ahz}
\begin{equation} \label{gamh}
\mathbf{\Gamma}_H = \sum_{i=1}^4 \Bigl[ \frac{C_i}{2} \Gamma_c (\alpha_s )\ln \frac{-t}{\mu^2} +
\gamma_i \Bigr] \mathbf{1}+\Gamma_c (\alpha_s) \mathbf{M}_H,
\end{equation}
where $\Gamma_c (\alpha_s)$ is the cusp anomalous dimension which can be 
expanded as~\cite{Korchemsky:1987wg}
\begin{equation}
\Gamma_c (\alpha_s) = \frac{\alpha_s}{4\pi} \Gamma_c^0 
+ \Bigl( \frac{\alpha_s}{4\pi}\Bigr)^2 \Gamma_c^1 +\cdots,
\end{equation}
with
\begin{equation}
\Gamma_c^0 =4, \  \Gamma_c^1 = \Bigl( \frac{268}{9}-\frac{4}{3}\pi^2 \Bigr) C_A 
-\frac{40 n_f}{9}.
\end{equation}  
To NLL accuracy, the cusp anomalous dimension to two loops is needed. The Casimir 
invariants $C_i$ are given by $C_q = C_F$,  $C_g = C_A$, and $\gamma_i$ are given by
\begin{equation}
\gamma_q = -\frac{\alpha_s}{2\pi} \frac{3}{2} C_F, \ \gamma_g = -\frac{\alpha_s}{2\pi} 
\frac{\beta_0}{2}.
\end{equation}
 
The matrix $\mathbf{M}_H$ can be written as
\begin{equation}
\mathbf{M}_H = -\sum_{i<j} \mathbf{T}_i \cdot \mathbf{T}_j \Bigl[L(s_{ij}) -L(t)\Bigr],
\end{equation}
where $s_{ij}$ are given as
\begin{align} \label{mandel}
&s=s_{12} = (p_1+ p_2)^2 = n_{12} \omega_1 \omega_2 = s_{34} = (p_3 + p_4)^2 = n_{34} 
\omega_3 \omega_4, \nonumber \\
&t= s_{13} = (p_1 -p_3)^2 = -n_{13} \omega_1 \omega_3 = s_{24} = (p_2 - p_4)^2 = -n_{24} 
\omega_2 \omega_4, \nonumber \\
& u= s_{14} = (p_1 -p_4)^2 = -n_{14} \omega_1 \omega_4 = s_{23} = (p_2 -p_3)^2 = -n_{23} 
\omega_2 \omega_3.
\end{align}
 And  
\begin{equation}
L(t) = \ln \frac{-t}{\mu^2}, \ L(u) = \ln \frac{-u}{\mu^2}, \ L(s) = \ln \frac{s}{\mu^2} -i\pi.
\end{equation}

Specifically, for $q\overline{q} \rightarrow gg$, the first part of $\mathbf{\Gamma}_H$ 
proportional to the identity matrix in Eq.~(\ref{gamh}) is written as
\begin{equation}
\frac{\alpha_s}{2\pi}\Bigl[ C_H   \ln \frac{n_{13}\omega_1 \omega_3}{\mu^2} 
 -3C_F -\beta_0 \Bigr] \mathbf{1} = \Bigl[\Gamma_c  \frac{C_H}{2} 
 \ln \frac{n_{13}\omega_1 \omega_3}{\mu^2} -2 \gamma_g -2\gamma_q\Bigr] \mathbf{1},
\end{equation}
where $C_H = n_q C_F + n_g C_A = 2 C_F + 2C_A$.
And the nondiagonal part $\mathbf{M}_H$ is given as
\begin{align} \label{mhmat}
 \mathbf{M}_H &=   \begin{pmatrix}
\displaystyle \frac{C_F}{2} \ln \frac{n_{12} n_{34}}{n_{13} n_{24}} & 0& \displaystyle 
 \ln \frac{n_{13} n_{24}}{n_{14} n_{23}} \\
0& \displaystyle   \frac{C_F}{2} \ln \frac{n_{12} n_{34}}{n_{13} n_{24}} 
+\frac{C_A}{2} \ln \frac{n_{14} n_{23}}{n_{13} n_{24}}& \displaystyle 
\ln \frac{n_{14} n_{23}}{n_{13} n_{24}} \\
\displaystyle  \frac{1}{4} \ln \frac{n_{12} n_{34}}{n_{14} n_{23}} & 
\displaystyle  \frac{1}{4} \ln \frac{n_{12} n_{34}}{n_{13} n_{24}} & 
\displaystyle  \frac{C_F+C_A}{2}  \ln \frac{n_{12} n_{34}}{n_{13} n_{24}}
\end{pmatrix}-i\pi \mathbf{T},
\end{align}
where $\mathbf{T}$ is given by
\begin{equation} \label{imagt}
\mathbf{T} =  \begin{pmatrix}
C_F&0&0\\
0&C_F&0\\
1/2&1/2& C_F +C_A
\end{pmatrix}.
\end{equation}
 
The imaginary part $-i\pi \mathbf{T}$ in $\mathbf{M}$ does not contribute to the evolution, 
and the solution to RG equation, Eq.~\eqref{hardrg}, is written as
\begin{equation}
\mathbf{H}(\mu_F) = \Pi_H (\mu_F, \mu_H) \bm{\Pi}_H (\mu_F,\mu_H)  \mathbf{H}(\mu_H)
\bm{\Pi}^{\dagger}_H (\mu_F, \mu_H), 
\end{equation}
where $\Pi_H$, and $\bm{\Pi}_H$ are given by
\begin{align} \label{hardk}
\Pi_H (\mu_F, \mu_H) &= \exp \Bigl[ -2 C_H S_{\Gamma} (\mu_F, \mu_H) - 
C_H a_{\Gamma} (\mu_F, \mu_H) \ln \frac{\mu_H^2}{-t} \nonumber \\
&+ 4 a_{\gamma_q} (\mu_F, \mu_H) 
+ 4 a_{\gamma_g} (\mu_F, \mu_H) \Bigr],
\nonumber \\
\bm{\Pi}_H (\mu_F, \mu_H) &= \exp \Bigl[ a_{\Gamma} (\mu_F, \mu_H) 
\mathbf{M}_H\Bigr]. 
\end{align}
Here $\Pi_H$ is the evolution kernal from the identity matrix in $\bm{\Gamma}_H$, and 
$\bm{\Pi}_H$ is the evolution from the matrix $\mathbf{M}_H$. 

The quantities in Eq.~\eqref{hardk} are defined as
\begin{equation}
S_{\Gamma} (\mu, \mu_i) = \int_{\alpha (\mu_i)}^{\alpha (\mu)} d\alpha 
\frac{\Gamma_{\mathrm{cusp}} (\alpha) }{\beta (\alpha)} \int_{\alpha (\mu_i)}^{\alpha} 
\frac{d\alpha'}{\beta(\alpha')},  \ \  \ a_f (\mu, \mu_i) = \int_{\alpha (\mu_i)}^{\alpha (\mu)}
\frac{d\alpha}{\beta(\alpha)} f(\alpha),
\end{equation}
where $f(\alpha)$ is a function of $\alpha$. The beta function $\beta(\alpha_s)$ is defined as
\begin{equation}
\beta(\alpha_s) = \mu \frac{d\alpha_s}{d\mu} = -2\alpha_s \Bigl[ \beta_0 
\Bigl( \frac{\alpha_s}{4\pi}\Bigr) + \beta_1 \Bigl( \frac{\alpha_s}{4\pi}\Bigr)^2+\cdots \Bigr],
\end{equation}
where
\begin{equation} \label{betaq}
\beta_0 = \frac{11}{3} C_A -\frac{4}{3}T_F n_f,  \ \ \beta_1 = \frac{34}{3} C_A^2 
- \frac{20}{3} C_A T_F n_f - 4C_F T_F n_f.
\end{equation}
 
\subsection{The beam function}
The RG equation for the beam function is written as
\begin{equation}
\frac{d}{d\ln \mu}  B_i (z_i, \delta_i \omega_i, \mu) = \gamma_{B_i} (\mu)  
B_i (z_i, \delta_i \omega_i,\mu),
\end{equation}
where the anomalous dimension for the beam function is given by
\begin{equation}
\gamma_{B_i} (\mu)  
= 2C_F \Gamma_c \ln \frac{\mu}{\delta_i \omega_i} -2\gamma_q.  
\end{equation}
The solution of the RG equation is written as
\begin{equation}
B_i (z_i, \delta_i \omega_i,\mu_F) = U_{B_i} (\mu_F, \mu_B) \mathcal{B}_i(z_i, \delta_i \omega_i,\mu_B),
\end{equation}
where the evolution kernel $U_B$ is given by
\begin{equation} \label{beamk}
U_{B_i} (\mu_F, \mu_B) = \exp \Bigl[ 2 C_F S_{\Gamma} (\mu_F, \mu_B) 
+ 2C_F a_{\Gamma} (\mu_F, \mu_B) \ln \frac{\mu_B}{\delta_i\omega_i} 
-2 a_{\gamma_q} (\mu_F, \mu_B)\Bigr].
\end{equation}

\subsection{The jet function}
The RG equation for the jet function is written as
\begin{equation}
\frac{d}{d\ln \mu} \mathcal{J}_{g_i} (\omega_i \delta_i, \mu) 
= \gamma_{J_i} \mathcal{J}_{g_i} (\omega_i \delta_i, \mu),
\end{equation}
where the anomalous dimension $\gamma_{J_i}$ is given by
\begin{equation}
\gamma_{J_i}   
= 2 C_A \Gamma_c \ln \frac{\mu}{\delta_i \omega_i} -2 \gamma_g. 
\end{equation}
The evolution of the jet function is given by
\begin{equation}
\mathcal{J}_{g_i} (\omega_i \delta_i, \mu_F) = U_{J_i} (\mu_F, \mu_J) 
\mathcal{J}_{g_i} (\omega_i \delta_i, \mu_J),
\end{equation}
where the evolution kernel $U_{J_i}$ is given as
\begin{equation} \label{jetk}
U_{J_i}  (\mu_F, \mu_J) = \exp \Bigl[ 2 C_A S_{\Gamma} (\mu_F, \mu_J) 
+ 2 C_A a_{\Gamma} (\mu_F, \mu_J) \ln \frac{\mu_J}{\delta_i \omega_i} 
-2 a_{\gamma_g} (\mu_F, \mu_J)\Bigr].
\end{equation}
 
\subsection{The soft function}
The soft function is renormalized in a matrix form as
\begin{equation}
\mathbf{S}_{\mathrm{bare}} = \mathbf{Z}_S^{\dagger} (\mu) \mathbf{S} (\mu) 
\mathbf{Z}_S (\mu),
\end{equation}
and the RG equation is given as
\begin{equation}
\frac{d}{d\ln \mu} \mathbf{S} = \mathbf{S} \mathbf{\Gamma}_S 
+\mathbf{\Gamma}_S^{\dagger} \mathbf{S},
\end{equation}
where the anomalous dimension matrix $\mathbf{\Gamma}_S$ is defined as
\begin{equation}
\mathbf{\Gamma}_S = - \Bigl(\frac{d}{d\ln \mu} \mathbf{Z}_S \Bigr) \mathbf{Z}_S^{-1}.
\end{equation}

The soft contribution $S_{ij}$ is given by Eq.~\eqref{totalsoft}, from which the 
anomalous soft dimensions can be obtained. Let us define $U_{ij}$ by 
\begin{equation}
\frac{\Gamma_c}{2} U_{ij} \equiv \frac{dS_{ij}}{d\ln \mu} = \Gamma_c 
\ln \frac{\delta_i \delta_j}{n_{ij}}.
\end{equation}
Then  the anomalous dimension matrix of the soft function is given as
\begin{equation} \label{gams}
\mathbf{\Gamma}_S = \Gamma_c \Bigl(-\frac{1}{2} \sum_{i<j} \mathbf{T}_i 
\cdot \mathbf{T}_j U^S_{ij} + i \pi \mathbf{T}\Bigr),
\end{equation}
 which can be written as
\begin{equation} \label{soanom}
\bm{\Gamma}_S =   \Bigl( C_F \Gamma_c \ln \frac{\delta_1 \delta_2}{n_{12}} + 
 C_A \Gamma_c  \ln \frac{\delta_3 \delta_4}{n_{34}} \Bigr) \mathbf{1} 
 +\Gamma_c (\mathbf{M}_S  + i\pi \mathbf{T}), 
\end{equation}
where $\mathbf{M}_S$ is given by
\begin{equation}  \label{msmat}
\mathbf{M}_S=\begin{pmatrix}
\displaystyle \frac{C_A}{2} \ln \frac{n_{12} n_{34}}{n_{13} n_{24}} & 0 & 
\displaystyle \ln \frac{n_{14} n_{23}}{n_{13} n_{24}} \\
 0 &\displaystyle \frac{C_A}{2} \ln \frac{n_{12} n_{34}}{n_{14} n_{23}} & 
 \displaystyle \ln \frac{n_{13} n_{24}}{n_{14} n_{23}}\\
\displaystyle \frac{1}{4} \ln \frac{n_{14} n_{23}}{ n_{12} n_{34}} & \displaystyle \frac{1}{4}
\ln \frac{n_{13} n_{24}}{n_{12} n_{34}}& 0
\end{pmatrix},
\end{equation}
and $\mathbf{T}$ is given by Eq.~\eqref{imagt}. We have inserted $\mathbf{T}$ to keep 
the consistency of the anomalous dimensions, and it does not affect Eq.~(\ref{sigmamu}) below
at one loop since $\mathbf{S}_0 \mathbf{T} = \mathbf{T}^{\dagger} \mathbf{S}_0$, where $\mathbf{S}_0$
is the tree-level soft function.

The evolution of the soft function from the scale $\mu_S$ to the common factorization scale $\mu_F$ 
can be written as
\begin{equation}
\mathbf{S} (\mu_F) = \Pi_S (\mu_F, \mu_S) \bm{\Pi}_S^{\dagger} (\mu_F, \mu_S) 
\mathbf{S} (\mu_S) \bm{\Pi}_S (\mu_F, \mu_S), 
\end{equation} 
where the evolution kernels are given as
\begin{align} \label{softk}
\Pi_S (\mu_F, \mu_S) &= \exp \Bigl[ 2C_F a_{\Gamma} (\mu_F, \mu_S) 
\ln \frac{\delta_1 \delta_2}{n_{12}} +2C_A a_{\Gamma}  (\mu_F, \mu_S) 
\ln \frac{\delta_3 \delta_4}{n_{34}} \Bigr], \nonumber \\
\bm{\Pi}_S (\mu_F, \mu_S) &=\exp \Bigl[ a_{\Gamma} (\mu_F, \mu_S) \mathbf{M}_S\Bigr].
\end{align}
The evolution kernel $\Pi_S$ is constructed from the diagonal part of the anomalous 
dimension matrix in Eq.~\eqref{soanom}, and $\bm{\Pi}_S$ from $\mathbf{M}_S$.

\subsection{Cancellation of the anomalous dimensions}
The dijet cross section should be independent of the renormalization scale, which means that 
the sum of the anomalous dimensions of the  factorized parts should be zero.  Then the dijet 
cross section is independent of the factorization scale $\mu_F$. At NLO, the derivative of 
the dijet cross section in Eq.~(\ref{nlsig}) with respect to $\ln \mu$ yields
\begin{align} \label{sigmamu}
\frac{d\sigma}{d\ln \mu} &= \Bigl[ \mathrm{tr} \ \mathbf{H}_0 \mathbf{S}_0 
\Bigl(\mathbf{\Gamma}_H +\mathbf{\Gamma}_S
+\frac{1}{2}(\gamma_{B_1} + \gamma_{B_2} +\gamma_{J_3} + \gamma_{J_4}) \mathbf{1}\Bigr)  
\nonumber \\
&+ \mathrm{tr} \ \mathbf{S}_0\mathbf{H}_0  \Bigl(\mathbf{\Gamma}_H^{\dagger}
 +\mathbf{\Gamma}_S^{\dagger} +\frac{1}{2}(\gamma_{B_1} + \gamma_{B_2} +\gamma_{J_3} 
 + \gamma_{J_4})\mathbf{1}\Bigr) \Bigr] \nonumber \\
&\times \Bigl( B_{q/N_1,0} B_{\bar{q}/N_2,0} + (q\leftrightarrow 
\overline{q})\Bigr)   \mathcal{J}_{g,0}   \mathcal{J}_{g,0}, 
\end{align}
where the quantities with the subscript 0 are the tree-level quantities.  The anomalous dimensions 
of the gluon jet function and the beam function are given by
\begin{equation}
\gamma_{B_i} = 2C_F \Gamma_c \ln \frac{\mu}{\delta_i \omega_i}-2 \gamma_q, \ \ 
\gamma_{J_i} =  2C_A \Gamma_c  \ln \frac{\mu}{\delta_i \omega_i} -2\gamma_g.
\end{equation}

From Eqs.~(\ref{gamh}), (\ref{gams}), the sum of the anomalous dimension matrices 
$\mathbf{\Gamma}_H+ \mathbf{\Gamma}_S$ becomes diagonal and is given by
\begin{align}
\mathbf{\Gamma}_H+ \mathbf{\Gamma}_S &=  \Gamma_c \Bigl[ \frac{C_A +C_F}{2} 
\ln \frac{n_{13}n_{24} \omega_1 \omega_2 \omega_3\omega_4}{\mu^4} 
+ \frac{C_A+C_F}{2} \ln \frac{n_{12} n_{34}}{n_{13} n_{24}} \nonumber \\
& + C_F \ln \frac{\delta_1 \delta_2}{n_{12}} +C_A \ln \frac{\delta_3 \delta_4}{n_{34}}\Bigr]  
 = -\frac{1}{2}\Bigl( \gamma_{B_1} + \gamma_{B_2} + \gamma_{J_3}+ \gamma_{J_4} \Bigr) \mathbf{1},
 \end{align}
 where the second relation obtained using Eq.~\eqref{mandel}.
Therefore the total sum of the anomalous dimensions is given by
\begin{equation}
 \mathbf{\Gamma}_H+ \mathbf{\Gamma}_S +\frac{1}{2}\Bigl( \gamma_{B_1} 
 + \gamma_{B_2} + \gamma_{J_3}+ \gamma_{J_4} \Bigr) \mathbf{1}=0.
\end{equation}
It is explicitly shown that the dijet cross section is independent of the renormalization scale $\mu$. 
Though the evolution kernels for each factorized part look complicated and depend on the 
factorization $\mu_F$ [See Eqs.~\eqref{hardk}, \eqref{beamk}, \eqref{jetk} and \eqref{softk}.], 
the product of all the evolution kernels turn out to be independent of $\mu_F$.
  
\section{Conclusion\label{conc}} 
The factorization theorems for various physical observables ranging from the jet cross sections 
to more differential quantities such as the jet substructure are crucial in theoretical prediction.  
Once the factorization theorems are established, each factorized part depends on a single scale 
and its evolution can be computed using perturbation theory. When there are hierarchies of 
scales, as often happens in high-energy scattering, large logarithms of the ratio of disparate 
scales appear and they can be resummed using the RG equation. However, in order to resum 
the large logarithms, we have to ensure that the divergences are purely of the UV origin. 
Therefore the absence of the IR and rapidity divergences in each factorized part is vital in 
proving factorization theorems. 

In this paper, we have considered the dijet cross section in hadron-hadron scattering by 
selecting a single partonic process $q \overline{q} \rightarrow gg$. We expect that the more
differential the physical quantities we probe, the more complicated the proof of the 
factorization becomes. And this is just the starting point in that direction. Though the 
dijet cross section looks simple, it contains a lot of interesting physics as we delved deeper 
into the divergence structure in this paper.

We should emphasize that there are three issues which are not included here. Firstly, we 
have considered only the case in which the jet cone size is not small. By dissecting the phase 
space into the collinear and the soft regions, each factorized part is described by a single 
scale. For example, the characteristic scales are $\mu_H \sim \omega$,
$\mu_J \sim \mu_B \sim \omega \delta$ and $\mu_S \sim \Lambda$ for the hard, collinear 
and soft scales respectively. The RG equation resums all the large logarithms when we scale 
from these scales to the factorization scale $\mu_F$. However, another large logarithm 
appears for small jet radius. The anomalous dimensions depend on the cone size, as can 
be seen in Eq.~\eqref{anomjet} for the jet function.  In order to resum the large logarithms 
for the small radius, we decompose the phase space more appropriately including the so-called 
soft-collinear modes~\cite{Dasgupta:2014yra,Chien:2015cka,Dasgupta:2016bnd}. The large 
logarithm due to the small radius can be handled, but in this paper, we take a reasonable 
size of the jet radius $R \sim 0.7$ such that the small-radius resummation gives a very small effect.

Secondly, in proving the factorization theorem, the interaction of the Glauber gluons between 
the active partons and the spectator partons should be considered to ensure the 
factorization~\cite{Bodwin:1984hc}. We assume that the Glauber gluons do not violate the 
factorization, as in many other processes. And finally the issue of nonglobal
logarithms should be raised, but if we choose appropriate observables such as the jettiness, 
the problem of the nonglobal logarithms can be avoided, though it should be considered 
at higher orders in the dijet cross section. In spite of these important issues, the factorization 
of the dijet cross section, the study of the divergence structure, and the resummation of 
large logarithms offer significant insights into the understanding of the hadron-hadron
scattering.
  
We have also computed the anomalous dimensions of the hard, collinear and soft parts. 
The anomalous dimensions of the collinear quark beam functions and the gluon jet functions are 
diagonal, and those of the hard and soft functions are nontrivial matrices in the operator basis. 
The jet algorithm and the beam veto affect both the diagonal gluon jet functions, the beam functions, 
and the non-diagonal soft function.  However, the dependence on the jet cone size in the soft 
function resides only in the diagonal part, and proportional to the identity matrix. This 
dependence of the jet cone radius in the soft function is cancelled by that in the beam and the 
jet functions. Also note that the nondiagonal part of the soft function in Eq.~\eqref{msmat} 
has only directional dependences of $n_{ij}$.  Interestingly enough, the nondiagonal part of 
the hard anomalous dimension matrix in Eq.~\eqref{mhmat} 
depends only on the directions, which cancel exactly that in the soft function. We have shown 
all these intertwined structure of the anomalous dimensions explicitly in the process
$q\overline{q} \rightarrow gg$. 
 
As already mentioned, the detailed analysis of extracting various divergences can be applied 
to other processes. It will be interesting to consider more differential observables probing 
jet substructure, such as angularity, $N$-jettiness, etc., and establish the factorization theorems 
and resum large logarithms. And as mentioned before, the issues of the small-$R$ resummation 
and the nonglobal logarithms will be pursued in the future.

\begin{acknowledgments}
The authors are supported by Basic Science Research Program through the National Research 
Foundation of Korea (NRF) funded by  the Ministry of Education (Grant No. NRF-2019R1F1A1060396),
and by BK21 FOUR program at Korea University, Initiative for Science Frontiers on Upcoming Challenges.  
\end{acknowledgments}

\bibliography{HadronJet.bib}
\end{document}